\documentclass{article}
\usepackage{arxiv}
\usepackage{latexsym}
\usepackage{graphicx}
\usepackage{amsfonts,amssymb,amsmath}
\usepackage{hyperref}
\usepackage{amsmath,amssymb}
\usepackage{tikz}
\usepackage{algorithm}
\usepackage{algorithmic}
\usepackage{pgfplots}
\usepackage{caption}
\usepackage{subcaption}
\usepackage{cite}
\usepackage{url}
\usetikzlibrary{calc}
\usepackage{amsthm}
 \usetikzlibrary{patterns}
\usepackage{multirow}
\usepackage{multicol}

\newtheorem{property}{Property}
\newtheorem{example}{Example}
\newtheorem{definition}{Definition}

\pgfplotsset{
    complexplot/.style={width=1.0\linewidth,xlabel near ticks,ylabel near ticks,grid=both,ylabel={$\Im m(.)$},xlabel={$\Re e(.)$},axis equal,xmin=-1.5,xmax=1.5,ymin=-1.5,ymax=1.5},
}

\renewcommand{\shorttitle}{PhyCOM: A Multi-Layer Parametric Network for JLIC and SD}

\author{ \href{https://orcid.org/0000-0002-2532-7486}{\includegraphics[scale=0.06]{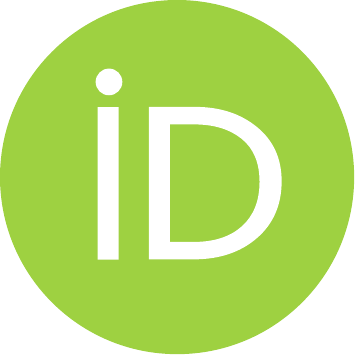}\hspace{1mm}Vincent Choqueuse}\\
	Lab-STICC, UMR CNRS 6285\\
	ENIB\\
	29238 Brest Cedex 3, France  \\
	\texttt{choqueuse@enib.fr} \\
	\And
	\href{https://orcid.org/0000-0001-6149-0195}{\includegraphics[scale=0.06]{orcid.pdf}\hspace{1mm}Alexandru Frunza} \\
	$^{(1)}$ Lab-STICC, UMR CNRS 6285\\
	ENIB\\
	29238 Brest Cedex 3, France  \\
	$^{(2)}$ Military Technical Academy\\
	Bulevardul George Cosbuc 39-49\\
	050141 Bucuresti, Romania\\
	\texttt{frunza@enib.fr} \\
	\And
	\href{https://orcid.org/0000-0001-8185-7134}{\includegraphics[scale=0.06]{orcid.pdf}\hspace{1mm}St\'ephane Azou} \\
	Lab-STICC, UMR CNRS 6285\\
	ENIB\\
	29238 Brest Cedex 3, France  \\
	\texttt{azou@enib.fr} \\
	\And
	\href{https://orcid.org/0000-0002-6671-1185}{\includegraphics[scale=0.06]{orcid.pdf}\hspace{1mm}Pascal Morel} \\
	Lab-STICC, UMR CNRS 6285\\
	ENIB\\
	29238 Brest Cedex 3, France  \\
	\texttt{morel@enib.fr} \\
}

\title{PhyCOM: A Multi-Layer Parametric Network for Joint Linear Impairments Compensation and Symbol Detection}

\begin{document}

\maketitle

\begin{abstract}
In this paper, we focus on the joint impairments compensation and symbol detection problem in communication systems. First, we introduce a new multi-layer channel model that represents the underlying physics of multiple impairments in communication systems. This model is composed of widely linear parametric layers that describe the input-output relationship of the front-end impairments and channel effects. Using this particular model, we show that the joint compensation and zero-forcing detection problem can be solved by a particular feedforward network called PhyCOM. Because of the small number of network parameters, a PhyCOM network can be trained efficiently using sophisticated optimization algorithms and a limited number of pilot symbols.

Numerical examples are provided to demonstrate the effectiveness of PhyCOM networks with communication systems corrupted by transmitter and receiver IQ imbalances, carrier frequency offset, finite impulse response channels, and transmitter and receiver phase noise distortions. Compared to conventional digital signal processing approaches, simulation results show that the proposed technique is much more flexible and offers better statistical performance both in terms of MSE and SER with a moderate increase of the computation complexity.\end{abstract}


\section{Introduction}

Future communication systems will integrate advanced technologies such as massive Multiple-Input Multiple-Output (MIMO), device-to-device communication, advanced coding schemes, and complex modulation formats to satisfy the increasing demand for high data rates~\cite{ELI15,BAS16,ZHA16,SHA17}. In practice, advanced coding techniques and modulation formats make the system's performance sensitive to signal distortion occurring at the physical layer~\cite{ELI15,WAN17}. These distortions are mainly caused by the propagation channel's effect and hardware imperfections, and include, for example, IQ imbalance, amplifier nonlinearities, phase noise, and carrier frequency offset (CFO). To avoid a severe degradation of the system performance, these distortions are usually mitigated with digital compensation algorithms at the receiver.

In the literature, many data-aided or blind algorithms have been developed for local impairment compensation. Such algorithms have been proposed to equalize the propagation channel~\cite{TUG00} and compensate IQ or carrier impairments~\cite{TAR05,TUB05,ANT08,SUN09,NAM12,YAO05,ZOU07}. Even if these local approaches can provide near-optimal statistical results for simple scenarios, these techniques can be difficult to use when  the communication system is corrupted by multiple impairments. When multiple impairments occur, state-of-the-art algorithms mainly use a global parametric approach. The global parametric approach assumes that the channel can be represented by a mathematical (or statistical) model that depends on several unknown parameters. Then, the compensation is usually obtained in two steps. First, the channel parameters are estimated globally using a trained, semi-blind, or blind technique~\cite{VAN97,TUG00,VAL01,GIL05,TAN07,TAR05,TAR07,TUB05,ANT08,HSU08,HOR08,INA09,CAI11,FRU21}. Secondly, the transmitted signal is detected from the received signal by replacing the channel by its estimate. The parametric approach is the preferred one when the statistical performance is of primary concern. Nevertheless, this approach usually leads to computationally demanding algorithms. Furthermore, this approach suffers from poor flexibility since changing the channel model usually implies new mathematical developments and leads to specific algorithms.

For joint channel estimation and symbol detection, a promising approach is based on machine learning. Machine learning techniques have shown to perform well in many engineering tasks, including image classification, speech and audio recognition, and natural language understanding~\cite{LEC15}. Recently, deep neural networks have been applied in physical layer communications to address detection problems~\cite{FAR18,SA19,HE19,KHA20,HE20,SHL20}, channel estimation and prediction~\cite{DEM19}, nonlinear fiber mitigation~\cite{HAG20}, or for the development of end-to-end communication systems~\cite{O2017,DOR17,SIM18}. Deep neural networks require little engineering by hand and have the distinct advantage of being more versatile than parametric approaches. Nevertheless, conventional networks are not always well adapted to communication problems occurring at the physical layer since they mainly rely on abstract mathematical models. To overcome this issue, several studies have demonstrated the advantage of combining both model-based approaches and deep neural networks. For the coherent MIMO detection problem, it has been shown that model-based deep learning techniques can achieve near maximum likelihood performance with lower computational complexity~\cite{SA19,KHA20,SHE20,HE20,SHL20}. Regarding the joint channel estimation and detection problem, a technique combining a non-parametric channel estimator and a model-based deep learning detector has been recently proposed in~\cite{HE20}. However, despite its generality, non-parametric channel estimators require a significant amount of pilot symbols when the channel matrix is large and so, are not well adapted for the compensation of fast time-varying impairments.

In this paper, we investigate the joint impairments compensation and symbol detection problem in communication systems using both parametric techniques and feedforward networks. As a preliminary study, we focus on the compensation of widely linear Single-Input Single-Output (SISO) impairments. For this problem, we show that the joint compensation and symbol detection problem can be modeled as a parametric MIMO detection problem. To address this problem, we propose a new network architecture called PhyCOM.

The main contributions of this paper are twofold.
\begin{itemize}
\item First, we introduce a new multi-layer model that represents the underlying physics of multiple widely linear impairments in communication systems. Using this model, we show that the joint compensation and zero-forcing detection can be solved using a general feedforward network, whatever the number and the type of widely linear impairments occurring at the physical layer and the pilot allocation strategy used. Compared to classical parametric compensation techniques such as the one recently described in~\cite{FRU21}, the proposed approach is more general and flexible since physical layers can be easily added, and their positions can be freely arranged in the network architecture. Contrary to non-parametric deep learning techniques that require a large amount of training data, simulations shows that the proposed network allows the tracking of fast time-varying impairments with a limited number of symbols (typically $50$ samples) and training iterations ($\approx 50$ iterations).
\item Secondly, we propose an efficient algorithm for network training based on a semi-supervised strategy and the Levenberg-Marquardt (LM) optimization algorithm. This algorithm exploits the parametric structure of physical impairments, the isomorphic property of particular layers, and the constellation of the transmitted symbols to reduce the computational complexity and improve the statistical performance of the training stage.  
\end{itemize}

The remainder of this paper is organized as follows. Section \ref{comchannel} presents the signal model. Section \ref{phycomarch} describes the PhyCOM network architecture and its associated training algorithm. Section \ref{commlayer} describes some commonly encountered SISO widely-linear impairment layers, and Section \ref{simu} reports on the performance of our proposed approach.\\

\textit{Notations:} In this paper, a lowercase boldface letter denotes a vector, and a capital boldface letter denotes a matrix. For any matrix $\mathbf{X}$, $\mathbf{X}^T$ and $\mathbf{X}^{-1}$ denote the transpose and the inverse, respectively. The matrix $\mathbf{I}_N$ is the $N\times N$ identity matrix and $\mathbf{1}_N$ is a $N\times 1$ all-ones vector. The symbols $\times$, $\otimes$, and $\odot$ correspond to the matrix product, Kronecker product, and Hadamard product, respectively. For any complex-valued vector $\mathbf{x}$, we introduce the augmented real-valued vector, $\tilde{\mathbf{x}}$, as follows
\begin{align}
\tilde{\mathbf{x}}&=\begin{bmatrix}
\Re e(\mathbf{x}) \\
\Im m(\mathbf{x})
\end{bmatrix}.
\end{align}
Similarly, the augmented matrix $\underline{\mathbf{X}}$ is defined as
\begin{align}
\underline{\mathbf{X}}&=\begin{bmatrix}
\Re e(\mathbf{X}) & -\Im m(\mathbf{X}) \\
\Im m(\mathbf{X}) &\Re e(\mathbf{X}) 
\end{bmatrix}.
\end{align}

\section{Signal Model}
\label{comchannel}

Let us consider an input data vector of length $N$ denoted by $\mathbf{s}=[s[0],\cdots,s[N-1]]^T$, where $s[n] \in \mathcal{S}$ and $\mathcal{S}$ is a finite alphabet composed of $|\mathcal{S}|$ complex elements (PSK, QAM). 
In a communication system, the input data vector is first transmitted to the analog front-end. Considering the channel's effect and the hardware transmitter and receiver impairments, the relationship between the transmitted and received signals can be described by a directed acyclic graph composed of $L$ layers, where each layer models a particular physical channel effect or hardware impairment.

\begin{figure}[!t]
\centering
\begin{tikzpicture}
\node[above] at (-1.5,1) {$\mathbf{s}$};
\draw[->,>=latex] (-1.5,1)--(-0.75,1);
\draw (-0.75,0.25) rectangle (0.75,1.75) node[pos=.5,text width=4em, text centered](l0) {\scriptsize $\Re e$ / $\Im m$ \\ Splitting} ;
\draw[->,>=latex] (0.75,1)--(1.75,1);
\node[above] at (1.25,1) {$\tilde{\mathbf{x}}_0$};
\draw (1.75,0.25) rectangle (3.25,1.75) node[pos=.5,text width=4em, text centered](l1) {Layer 1} ;
\draw[->,>=latex] (3.25,1)--(4.25,1);
\draw[dashed] (4.25,0.25) rectangle (5.75,1.75) node[pos=.5,text width=4em, text centered] {$\cdots$};
\draw[->,>=latex] (5.75,1)--(6.75,1);
\draw (6.75,0.25) rectangle (8.25,1.75) node[pos=.5,text width=4em, text centered]  (l3){Layer $L$};
\draw[->,>=latex] (8.25,1)--(9,1); 
\node[above] at (8.75,1) {$\tilde{\mathbf{x}}_L$};
\end{tikzpicture}
\caption{Multi-Layer Physical Model. The received signal is obtained by propagating the augmented input signal into a feedforward network composed of widely linear layers.}\label{sys1}
\end{figure}
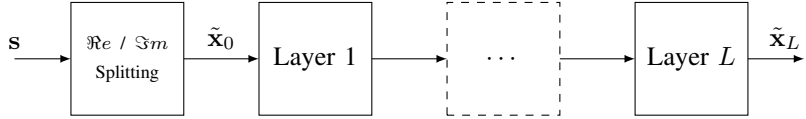

For illustration purposes, Fig.~\ref{sys1} illustrates a communication chain composed of $L$ widely linear layers. The first layer converts the complex vector $\mathbf{s}$ into its augmented version $\tilde{\mathbf{x}}_0=[\Re e(\mathbf{s}^T),\Im m(\mathbf{s}^T)]^T \in \mathcal{M}^{2N}$, where $ \mathcal{M}$ denotes the constellation of each element of $\tilde{\mathbf{x}}_0$. Then, the remaining layers transform an input augmented vector into an output augmented vector having the same dimension. 

\subsection{Widely Linear Layers}

As a preliminary study, this paper focuses on SISO widely linear layers. For widely linear layers, the output of the $l^{th}$ layer can be modeled as
\begin{align}
\tilde{\mathbf{x}}_l = \mathbf{F}_l(\boldsymbol \alpha_l)\tilde{\mathbf{x}}_{l-1}\label{eqmod1},
\end{align}
where 
\begin{itemize}
\item $\tilde{\mathbf{x}}_{l-1}\in \mathbb{R}^{2N}$ and $\tilde{\mathbf{x}}_{l}\in \mathbb{R}^{2N}$ are the real-valued augmented input and output of the $l^{th}$ layer
\item $\mathbf{F}_l(\boldsymbol  \alpha_l)$ is the $2N\times 2N$ square transfer matrix whose structure depends on the physical properties of the $l^{th}$ layer,
\item $\boldsymbol  \alpha_l=\begin{bmatrix}\alpha_l[1],\cdots, \alpha_l[K_l]\end{bmatrix}^T$ is a vector containing the layer parameters.
\end{itemize}

As illustrated in Section~\ref{commlayer}, widely linear layers can model a wide range of physical impairments such as IQ imbalance, CFO, Phase Noise, and Finite Impulse Response (FIR) channels. More restrictive models are also commonly encountered, such as strictly linear and isomorphic models.

\begin{definition}[Strictly Linear Layer]
For strictly linear layers, the transfer matrix can be decomposed as
\begin{align}
\mathbf{F}_l(\boldsymbol \alpha)=\underline{\mathbf{M}}_l(\boldsymbol \alpha)=\begin{bmatrix}
\Re e(\mathbf{M}_l(\boldsymbol \alpha)) & -\Im m(\mathbf{M}_l(\boldsymbol \alpha)) \\
\Im m(\mathbf{M}_l(\boldsymbol \alpha)) &\Re e(\mathbf{M}_l(\boldsymbol \alpha)) 
\end{bmatrix}.
\end{align}
where $\mathbf{M}_l(\boldsymbol \alpha_l)$ is an $N\times N$ complex-valued matrix.
\end{definition}

\begin{definition}[Isomorphic Layer]\label{defiso}
For isomorphic layers, the inverse of the transfer matrix can be expressed as
\begin{align}
\mathbf{F}_l(\boldsymbol \beta)=\mathbf{F}^{-1}_l(\boldsymbol \alpha),\label{refiso}
\end{align}
where $\boldsymbol \beta=\mathbf{g}(\boldsymbol \alpha)$ is a $K_l \to K_l$ reverse propagation function. 
\end{definition}

In the next section, we show that the isomorphic property allows the use of very low-complexity algorithms for layer compensation since it avoids a matrix inversion.

\subsection{Channel Model}

Mathematically, the channel output is given by the augmented vector $\tilde{\mathbf{x}}_L$ (see Fig.~\ref{sys1}). By also including a noise component, the received output can be finally modeled as
\begin{align}
\tilde{\mathbf{y}}_0 = \tilde{\mathbf{x}}_L +\tilde{\mathbf{b}},
\end{align}
where $\tilde{\mathbf{b}}$ corresponds to the augmented noise contribution, and
\begin{align}
\tilde{\mathbf{x}}_L = \mathbf{F}_{tot}(\boldsymbol \alpha)\tilde{\mathbf{x}}_{0},\label{eq_comp0}
\end{align}
with
\begin{itemize}
\item $\mathbf{F}_{tot}(\boldsymbol  \alpha)$ corresponds to the accumulated $2N\times 2N$ transfer channel matrix that is defined by
\begin{align}
\mathbf{F}_{tot}(\boldsymbol  \alpha)=\mathbf{F}_{L}(\boldsymbol  \alpha_L)\times \cdots \times \mathbf{F}_{1}(\boldsymbol  \alpha_1),
\end{align}
\item $\tilde{\mathbf{x}}_{0} \in \mathcal{M}^{2N}$ corresponds to the augmented input vector,
\item $\tilde{\mathbf{x}}_{L} \in \mathbb{R}^{2N}$ corresponds to the augmented output vector,  
\item $\boldsymbol\alpha = [\boldsymbol \alpha_{1}^T,\cdots,\boldsymbol \alpha_{L}^T]^T$ is a column vector containing the $K = \sum_{l=1}^L K_l$ real-valued channel parameters.
\end{itemize}

In practice, the statistical distribution of the noise component $\tilde{\mathbf{b}}$ usually depends on the value of the channel parameters. For ease of simplicity, this study neglects the dependence of the layer parameters $\boldsymbol  \alpha$ on the noise distribution.

\section{PhyCOM Network}
\label{phycomarch}

At the receiver side, the objective is to detect the vector $\mathbf{s}$, or its augmented version $\tilde{\mathbf{x}}_{0}$, from the received samples $\tilde{\mathbf{y}}_0 $. The detection can be achieved by minimizing a cost function $\mathcal{L}(.)$ as follows
\begin{align}
\widehat{\tilde{\mathbf{x}}}_{0}=\arg \min_{\tilde{\mathbf{x}}_{0}\in \mathcal{M}^{2N}} \mathcal{L}\left(\tilde{\mathbf{y}}_0, \mathbf{F}_{tot}(\boldsymbol \alpha)\tilde{\mathbf{x}}_{0}\right).
\end{align}
This problem corresponds to a parametric MIMO detection problem and it is challenging for two reasons. First, the minimization involves an exhaustive search over the $|\mathcal{M}|^{2N}$ symbol combinations. Secondly, the estimation of $\boldsymbol \alpha$ and $\tilde{\mathbf{x}}_{0}$ must be performed jointly since the channel parameters are usually unknown at the receiver side.

\subsection{Related Problems}

When the parametric structure of the $2N\times 2N$ matrix $\mathbf{F}_{tot}(\boldsymbol \alpha)$ is relaxed, the optimization problem is equivalent to a MIMO detection problem. In a MIMO detection problem, the objective is to minimize the cost function $\mathcal{L}\left(\tilde{\mathbf{y}}_0, \mathbf{F}\tilde{\mathbf{x}}_{0}\right)$ with respect to $\tilde{\mathbf{x}}_{0}\in \mathcal{M}^{2N}$.

\begin{itemize}
\item When $\mathbf{F}$ is known, the minimization problem simplifies to a coherent detection problem \cite{LAR09}. In the last decades, several coherent detectors have been proposed such as the Maximum Likelihood technique, Zero-Forcing (ZF) detector, Minimum Mean Square Error detector (MMSE), and the Sphere Decoder. Among them, the Maximum Likelihood detector gives the best statistical performance but suffers from high computational complexity. Recently, several alternatives based on deep-learning networks have been proposed \cite{SA19,KHA20,SHL20}. In particular, it has been shown that the DetNet detector proposed in \cite{SA19} can lead to near ML statistical performance with a significant reduction of the computational complexity.
\item When $\mathbf{F}$ is unknown, the MIMO detection problem is more challenging since it requires minimizing $\mathcal{L}\left(\tilde{\mathbf{y}}_0, \mathbf{F}\tilde{\mathbf{x}}_{0}\right)$ with respect to $\tilde{\mathbf{x}}_{0}$ and $\mathbf{F}$. In the literature, several neural network techniques have been proposed that address the detection problem without requiring the channel matrix as a prerequisite~\cite{FAR18,SA19}. Nevertheless, these black-box detectors require a huge training database and cannot capture the dependencies of changing channels~\cite{SA19}. To overcome this problem, a conventional approach is to use a training block of symbols to estimate $\mathbf{F}$ and then use this channel estimate with a coherent detector. Classical channel estimation techniques include the Least Squares (LS) estimators and the minimum Mean Square-Error Estimators (MMSE)~\cite{BIG06}. An architecture that combines both MMSE channel estimation and deep-learning detection has been recently proposed in \cite{HE20}. This architecture outperforms conventional approaches by compensating for residual channel estimation errors when performing symbol detection.
\end{itemize}

For the challenging case where $\mathbf{F}$ is unknown, most existing algorithms are composed of a channel estimation step. When the parametric structure of the $2N\times 2N$ matrix $\mathbf{F}_{tot}(\boldsymbol \alpha)$ is relaxed, conventional trained-based channel estimators require at least $2N$ blocks of training data~\cite{BIG06}, i.e., a total number of $(2N)^2$ training symbols. This requirement can be problematic with time-varying impairments, i.e., when some elements of $\boldsymbol \alpha$ vary between two blocks of data and when the number of known pilot symbols is limited. To address this issue, this study proposes a joint parameter estimation and data detection algorithm that exploits the multi-layer parametric model of $\mathbf{F}_{tot}(\boldsymbol \alpha)$. As stated in~\cite{CHE17}, exploiting the structure of a parameterized matrix $\mathbf{F}_{tot}(\boldsymbol \alpha)$ allows to dramatically accelerate the inference and training stage~\cite{CHE17}. Furthermore, the use of a parameterized channel model also allows to detect the payload data from few pilots.

\subsection{Model-Based Compensation Network}

\subsubsection{Motivations}
The architecture of the PhyCOM network is derived from the clairvoyant detector, which assumes knowledge of the accumulated transfer matrix $\mathbf{F}_{tot}(\boldsymbol \alpha)$. Note that this assumption is explicitly relaxed in the next subsection.

For the clairvoyant detector, the transmitted complex symbols $\mathbf{x}_{0}$ can be detected using a ZF detector. The ZF detector is a linear detector that is composed of two steps. 
\begin{itemize}
\item First, this detector estimates the transmitted samples by relaxing the finite alphabet constraint as follows
\begin{align}
\widehat{\tilde{\mathbf{x}}}_{0}^{ZF}=\arg \min_{\tilde{\mathbf{x}}_{0}\in \mathbb{R}^{2N}} \| \tilde{\mathbf{y}}_{0}- \mathbf{F}_{tot}(\boldsymbol \alpha)\tilde{\mathbf{x}}_{0}\|^2.
\end{align}
As $ \mathbf{F}_{tot}(\boldsymbol \alpha)$ is a square matrix, the solution of this unconstrained estimation problem is simply given by
\begin{align}
\widehat{\tilde{\mathbf{x}}}_{0}^{ZF}&=\mathbf{F}^{-1}_{tot}(\boldsymbol \alpha) \tilde{\mathbf{y}}_{0}=\mathbf{F}_{1}^{-1}(\boldsymbol  \alpha_1)\times \cdots \times \mathbf{F}_{L}^{-1}(\boldsymbol  \alpha_L)\tilde{\mathbf{y}}_{0}
\label{eq_comp1}.
\end{align}
\item Secondly, the transmitted symbols are detected by projecting each element of $\widehat{\tilde{\mathbf{x}}}_{0}^{ZF}$ into the constellation set $ \mathcal{M}$ as follows $\widehat{\tilde{\mathbf{x}}}_{0}=\boldsymbol{\Pi}_{\mathcal{M}}(\widehat{\tilde{\mathbf{x}}}_{0}^{ZF})$.
\end{itemize}

\begin{figure}[!t]
\centering
\begin{tikzpicture}
\node[above] at (-1.5,1) {$\tilde{\mathbf{y}}_0$};
\draw[->,>=latex] (-1.5,1)--(-0.75,1);
\draw (-0.75,0.25) rectangle (0.75,1.75) node[pos=.5,text width=4em, text centered](l0) {Layer $1$} ;
\draw[->,>=latex] (0.75,1)--(1.75,1);
\node[above] at (1.25,1) {$\tilde{\mathbf{y}}_1$};
\draw[dashed]  (1.75,0.25) rectangle (3.25,1.75) node[pos=.5,text width=4em, text centered](l1) {$\cdots$} ;
\draw[->,>=latex] (3.25,1)--(4.25,1);
\draw(4.25,0.25) rectangle (5.75,1.75) node[pos=.5,text width=4em, text centered] {Layer $L$};
\draw[->,>=latex] (5.75,1)--(6.75,1);
\node[above] at (6.25,1) {$\tilde{\mathbf{y}}_L$};
\draw (6.75,0.25) rectangle (8.25,1.75) node[pos=.5,text width=4em, text centered]  (l3){\scriptsize Non-Linear Detection};
\draw[->,>=latex] (8.25,1)--(9,1); 
\node[above] at (9,1) {$\widehat{\mathbf{s}}$};
\end{tikzpicture}
\caption{PhyCOM network architecture.}\label{PhyCOM_fig}
\end{figure}
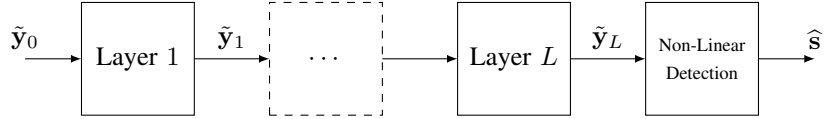

The expression of the ZF detector in~\eqref{eq_comp1} suggests a simple network architecture for impairments compensation and symbol detection. This architecture is presented in Fig.~\ref{PhyCOM_fig} and is composed of $L$ learnable compensation layers followed by a non-linear detection layer.

%

\subsubsection{Compensation layers} 

In Fig.~\ref{PhyCOM_fig}, the output of the $l^{th}$ compensation layer can be expressed as follows
\begin{align}
\tilde{\mathbf{y}}_{l} =\mathbf{H}_{l}(\boldsymbol \theta_l) \tilde{\mathbf{y}}_{l-1}.
\end{align}
where $ \mathbf{H}_l(\boldsymbol  \theta_l)$ is the layer $2N\times 2N$ transfer matrix. The output of the $L$ compensation layers is then given by
\begin{align}
\tilde{\mathbf{y}}_{L}&=\mathbf{H}_L(\boldsymbol  \theta_L)\times \cdots \times \mathbf{H}_{1}(\boldsymbol  \theta_1)\tilde{\mathbf{y}}_{0}\label{eq_comp2}.
\end{align}

Note that the performance of the PhyCOM network critically depends on the choice of the transfer matrices $\mathbf{H}_l(\boldsymbol  \theta_l)$. The next property describes a simple choice that guarantee $\tilde{\mathbf{y}}_{L}=\tilde{\mathbf{x}}_{0}$ under noiseless conditions.

\begin{property}[Noiseless conditions]\label{propnoiseless}
Under noiseless conditions, the impairments can be fully compensated by setting
\begin{align}
 \mathbf{H}_{l}(\boldsymbol \theta_l) = \mathbf{F}_{L-l+1}^{-1}(\boldsymbol \theta_l).
\end{align}
with $\boldsymbol \theta_l=\boldsymbol \alpha_{L-l+1}$.
\end{property}
\begin{proof}
Under noiseless conditions, the equations \eqref{eq_comp0} and \eqref{eq_comp2} show that
\begin{align*}
\tilde{\mathbf{y}}_{L}&=\mathbf{H}_L(\boldsymbol  \theta_L)\times \cdots \times \mathbf{H}_{1}(\boldsymbol  \theta_1) \mathbf{F}_{tot}(\boldsymbol \alpha)\tilde{\mathbf{x}}_{0}\\
&= \mathbf{H}_L(\boldsymbol  \theta_L)\times \cdots \times \mathbf{H}_{1}(\boldsymbol  \theta_1)\mathbf{F}_{L}(\boldsymbol  \alpha_L)\times \cdots \times \mathbf{F}_{1}(\boldsymbol  \alpha_1)\tilde{\mathbf{x}}_{0}
\end{align*}
By setting $ \mathbf{H}_{l}(\boldsymbol \theta_l) = \mathbf{F}_{L-l+1}^{-1}(\boldsymbol \alpha_{L-l+1})$, we obtain
\begin{align*}
\tilde{\mathbf{y}}_{L}&= \mathbf{F}_{1}^{-1}(\boldsymbol  \theta_1)\times \cdots \times \mathbf{F}_{L}^{-1}(\boldsymbol  \theta_L)\mathbf{F}_{L}(\boldsymbol  \alpha_L)\times \cdots \times \mathbf{F}_{1}(\boldsymbol  \alpha_1)\tilde{\mathbf{x}}_{0}=\tilde{\mathbf{x}}_{0}
\end{align*}
\end{proof}

In practice, the evaluation of the transfer matrix $ \mathbf{H}_{l}(\boldsymbol \theta_l)$ requires the inversion of a $2N\times 2N$ real-valued matrix. This inversion can be avoided for isomorphic layers using the \emph{Isomorphic Trick}.

\begin{property}[Isomorphic Trick] If the $l^{th}$ compensation layer satisfies the isomorphic equality in \eqref{refiso}, the transfer matrix of the $l^{th}$ compensation layer can be expressed as
\begin{align}
 \mathbf{H}_{l}(\boldsymbol \theta_l) = \mathbf{F}_{L-l+1}(\mathbf{g}(\boldsymbol \alpha_{L-l+1})).
\end{align}
\end{property}
\begin{proof}
The proof comes directly from Definition \ref{propnoiseless} and Property \ref{defiso}. 
\end{proof}

\subsubsection{Non-Linear Detection Layer}

The non-linear detection layer performs two operations. 

\begin{itemize}
\item Euclidean projection. First, each element of $\tilde{\mathbf{y}}_{L}$ is projected into $\mathcal{M}$ using the $\mathcal{L}_2$-norm, i.e., 
\begin{align}
\widehat{\tilde{\mathbf{x}}}_0=\boldsymbol{\Pi}_{\mathcal{M}}(\tilde{\mathbf{y}}_{L})\triangleq \arg \min_{\tilde{\mathbf{x}} \in \mathcal{M}^{2N}} \|\tilde{\mathbf{x}} -\tilde{\mathbf{y}}_{L} \|^2_2.
\end{align}
The solution of this problem can be obtained by performing $2N$ independent Euclidean projections. More precisely, the $n^{th}$ element of $\widehat{\tilde{\mathbf{x}}}_0=[\widehat{\tilde{x}}_{0}[0],\cdots,\widehat{\tilde{x}}_{0}[2N-1]]$ can be computed as
\begin{align}
\widehat{\tilde{x}}_{0}[n]=\arg \min_{\tilde{x}\in \mathcal{M}}|\tilde{x}-\tilde{y}_{L}[n]|^2_2.
\end{align}
\item Real to Complex transformation. Secondly, the complex vector $\widehat{\mathbf{s}}$ is constructed from the real-valued vector $\widehat{\tilde{\mathbf{x}}}_0$ using
\begin{align}
\widehat{\mathbf{s}}=\begin{bmatrix}\mathbf{I}_N&j\mathbf{I}_N\end{bmatrix}\widehat{\tilde{\mathbf{x}}}_0.
\end{align}
\end{itemize}


%

\subsection{Network Training}
\label{sec_nettraining}

In practice, the parameters $\boldsymbol \theta$ are unknown at the receiver side and must be estimated during a training stage. In this section, we describe a semi-supervised strategy for parameters estimation. This strategy is composed of a supervised stage followed by a self learning stage.

\subsubsection{Pilot Allocation}

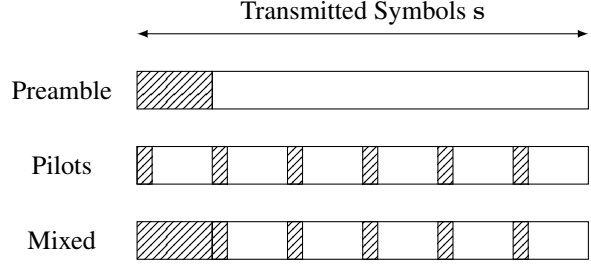
\begin{figure}[!t]
\centering
\begin{tikzpicture}[]
\draw[<->,>=latex] (0,1) -- node[above](){Transmitted Symbols $\mathbf{s}$} ( 6,1);
\draw[pattern=north east lines, pattern color=black] (0,0) rectangle (1,0.5) node[pos=.5,text width=4em, text centered] {} ;
\draw (1,0) rectangle (6,0.5) node[pos=.5,text width=6em, text centered] {} ;
\node at (-1,0.25) {Preamble};
\foreach \x in {0,...,5} 
       {\draw[pattern=north east lines, pattern color=black]  (1*\x,-1) rectangle (\x+0.2,-0.5);}
       
\draw (0,-1) rectangle (6,-0.5) node[pos=.5,text width=4em, text centered] {} ;
\node at (-1,-0.75) {Pilots};
\foreach \x in {1,...,5} 
       {\draw[pattern=north east lines, pattern color=black]  (1*\x,-2) rectangle (\x+0.2,-1.5);}
\draw (1,-2) rectangle (6,-1.5) node[pos=.5,text width=6em, text centered] {} ;
\draw[pattern=north east lines, pattern color=black] (0,-2) rectangle (1,-1.5) node[pos=.5,text width=4em, text centered] {};
\node at (-1,-1.75) {Mixed};
\end{tikzpicture}
\caption{Training symbols allocation strategies. The dash rectangles indicate the position of the $N_p$ training symbols $\mathbf{s}_{P}=\mathbf{P}\mathbf{s}$.}\label{fig_allocate}
\end{figure}

During the supervised stage, we assume that $N_p\ll N$ pilot symbols are known at the receiver. We also assume that the pilot symbols, $\mathbf{s}_{P}$, can be extracted from the input data, $\mathbf{s}$, using a matrix multiplication as follows
\begin{align}
\mathbf{s}_{P}&=\mathbf{P}\mathbf{s},
\end{align}
where $\mathbf{P}$ is an $N_p\times N$ \emph{allocation matrix}. An allocation matrix is a binary matrix that contains a single $1$ in each row and zeroes elsewhere. Some commonly used strategies for pilot allocation are illustrated in Fig.~\ref{fig_allocate}. The expression of the allocation matrix, $\mathbf{P}$, for preamble-based and pilot-based allocations are also described below.

\begin{example}[Preamble-based allocation]
In preamble-based allocation, a sequence of $N_p$ consecutive training symbols is followed by a block of data. The corresponding allocation matrix is given by 
$$\mathbf{P}=\begin{bmatrix}
\mathbf{I}_{N_p} & \mathbf{0}_{N_P\times (N-N_p)}
\end{bmatrix}.$$
\end{example}

\begin{example}[Pilot-based allocation]
In pilot-based allocation, $N_p$ pilots are sent periodically. The corresponding allocation matrix is given by 
$$\mathbf{P}= \mathbf{I}_{N_p}\otimes \begin{bmatrix}
1 & 0 & \cdots & 0
\end{bmatrix}.$$
\end{example}

\subsubsection{Supervised Training}

Let us denote the initial network parameters by $\boldsymbol\theta = [\boldsymbol \theta_{1}^T,\cdots,\boldsymbol \theta_{L}^T]^T$. To estimate the network parameters, we propose to minimize the least-squares error between the $N_P$ pilot symbols and the network output using an iterative optimization algorithm. Nevertheless, instead of using the output of the non-linear detection layer, which is non-differentiable, we propose to minimize the \emph{pre-detection} error between the augmented pilot symbols, $\tilde{\mathbf{x}}_{0P}=\left(\mathbf{I}_2\otimes \mathbf{P}\right)\tilde{\mathbf{x}}_{0}$, and the input of the non-linear detection layer, $\left(\mathbf{I}_2\otimes \mathbf{P}\right)\tilde{\mathbf{y}}_{L}$. Mathematically, this optimization problem is given by
\begin{align}
\widehat{\boldsymbol \theta}=\arg \min_{\boldsymbol \theta}\frac{1}{2}\|\mathbf{f}(\boldsymbol \theta)\|^2_2,\label{eqcost}
\end{align}
where 
\begin{align}
\mathbf{f}(\boldsymbol \theta)&=\tilde{\mathbf{x}}_{0P}-\left(\mathbf{I}_2\otimes \mathbf{P}\right)\tilde{\mathbf{y}}_{L},\\
\tilde{\mathbf{y}}_{L}&=\mathbf{H}_{L}(\boldsymbol  \theta_L)\times \cdots \times \mathbf{H}_{1}(\boldsymbol  \theta_1)\tilde{\mathbf{y}}_{0}.
\end{align}

To address the minimization problem in \eqref{eqcost}, we propose to use the following iterative procedure 
\begin{align}
\boldsymbol \theta \leftarrow \boldsymbol \theta +\mathbf{h},\label{eqgrad}
\end{align}
where $\mathbf{h}$ corresponds to a directional vector. In classical deep learning architecture, the directional vector is usually updated using stochastic gradient descent. 
As PhyCOM network usually depends on a smaller number of parameters, the network can be trained efficiently using more sophisticated optimization algorithms such as the Levenberg-Marquardt (LM) method. This optimization method allows to reduce the training time drastically~\cite{NOC06,YU11}.
For example, while a conventional Gradient Descent typically requires more 10000 iterations to train the PhyCOM network described in Fig.~\ref{wirelesschannel}, the LM method usually requires less than $20$ iterations.

Using the LM method, the directional vector is given by
\begin{align}
\mathbf{h}&=-(\mathbf{J}^T(\boldsymbol \theta)\mathbf{J}(\boldsymbol \theta)+\mu \mathbf{I})^{-1}\mathbf{J}^T (\boldsymbol \theta)\mathbf{f}(\boldsymbol \theta),\label{eqgrad2}
\end{align}
where $\mathbf{J}(\boldsymbol \theta)$ is the $2N\times K$ Jacobian matrix, and $\mu$ is a step size. The Jacobian matrix can be decomposed in $L$ blocks as follows 
\begin{align}
\mathbf{J}(\boldsymbol \theta) \triangleq 
\begin{bmatrix}
\mathbf{J}_1(\boldsymbol \theta)&\cdots&\mathbf{J}_L(\boldsymbol \theta)
\end{bmatrix},\label{eqJTOT}
\end{align}
where the $2N\times K_l$ matrix $ \mathbf{J}_l(\boldsymbol \theta)$ corresponds to the \emph{partial Jacobian} for the $l^{th}$ layer. Using the expression of $\mathbf{f}(\boldsymbol \theta)$, this partial Jacobian can be expressed as
\begin{align}
\mathbf{J}_l(\boldsymbol \theta)=-\left(\mathbf{I}_2\otimes \mathbf{P}\right) \times \mathbf{H}_{L}(\boldsymbol \theta_L)\times \cdots \times \mathbf{H}_{l+1}(\boldsymbol \theta_{l+1})\times \mathbf{L}_l(\boldsymbol \theta_l),
\label{eqJl}
\end{align}
where $\mathbf{L}_l(\boldsymbol \theta_l)$ corresponds to the \emph{local Jacobian} of the $l^{th}$ layer. This local Jacobian is a $2N\times K_l$ matrix defined as
\begin{align}
\mathbf{L}_l(\boldsymbol \theta_l)&=\left[
\frac{\partial \tilde{\mathbf{y}}_{l}}{\partial  \theta_l[1]}\cdots\frac{\partial \tilde{\mathbf{y}}_{l}}{\partial \theta_l[K_l]}
\right]\nonumber\\
&= \begin{bmatrix}
\frac{\partial \mathbf{H}_{l}(\boldsymbol \theta_{l}) }{\partial  \theta_l[1]}\tilde{\mathbf{y}}_{l-1}&\cdots&\frac{\partial \mathbf{H}_{l}(\boldsymbol \theta_{l}) }{\partial  \theta_l[K]}\tilde{\mathbf{y}}_{l-1}
\end{bmatrix}\label{eqlocjacob},
\end{align}
where $\theta_l[k]$ is the $k^{th}$ parameter of the $l^{th}$ layer.

In practice, the most computationally demanding task is evaluating the Jacobian matrix. Instead of using a naive approach, we propose to evaluate the $L$ partial Jacobians efficiently using a backpropagation strategy that mimics the principle of backpropagation algorithms in neural networks. Let us define the initial $2N_p\times 2N$ backpropagation matrix as 
\begin{align}
\mathbf{B}_{L+1}&=-\left(\mathbf{I}_2\otimes \mathbf{P}\right).
\end{align}
Using the definition of $\mathbf{B}_{L+1}$, the partial Jacobian matrix for the layer $l$ can be computed as 
\begin{align}
\mathbf{J}_l(\boldsymbol \theta)= \mathbf{B}_{l+1}\mathbf{L}_l(\boldsymbol \theta_l),
\end{align}
where $\mathbf{B}_{L+1}$ is backpropagated from layer $L$ to layer $1$ using the  iterative procedure
\begin{align}
\mathbf{B}_{l} =\mathbf{B}_{l+1}  \mathbf{H}_{l}(\boldsymbol \theta_{l}).
\end{align}
Finally, the proposed strategy for supervised network training is summarized in Algorithm \ref{algo1}. This algorithm is based on the efficient and robust implementation of the LM method described in~\cite{MORE78}. In particular, it is worth mentioning that this implementation does not require to evaluate the Jacobian matrix in each iteration.


%

%

 \begin{algorithm}[!t]
 \caption{PhyCOM : Supervised Learning}
 \begin{algorithmic}[1]
 \renewcommand{\algorithmicrequire}{\textbf{Input:}}
 \renewcommand{\algorithmicensure}{\textbf{Output:}}
 \REQUIRE training data $\{\tilde{\mathbf{y}}_0,\tilde{\mathbf{x}}_{0P}\}$, allocation matrix $\mathbf{P}$. 
 \STATE Initialize $\boldsymbol \theta$.
  \WHILE{a stopping criteria is not met}
  \STATE  //FORWARD PASS
  \FOR{l=$1$ to $L$}
  	 \STATE Compute and store the matrix $\mathbf{H}_{l}(\boldsymbol \theta_{l})$.
   	\STATE Compute the layer outputs as $\tilde{\mathbf{y}}_{l}=\mathbf{H}_{l}(\boldsymbol \theta_l) \tilde{\mathbf{y}}_{l-1}$.
 \ENDFOR
 \STATE Evaluate $\mathbf{f}(\boldsymbol \theta)=\tilde{\mathbf{x}}_{0P}-\left(\mathbf{I}_2\otimes\mathbf{P}\right)\tilde{\mathbf{y}}_L$.
\IF{required\_jacobian=True} 
  \STATE  //BACKWARD PASS
  \STATE Initialize $\mathbf{B}_{L+1}=-\mathbf{I}_2\otimes \mathbf{P}$.
 \FOR{l=$L$ to $1$}
  \STATE Compute the local Jacobian $ \mathbf{L}_l(\boldsymbol \theta_{l})$ using \eqref{eqlocjacob}.
   \STATE $\mathbf{J}_l(\boldsymbol \theta_{l})=\mathbf{B}_{l+1}\mathbf{L}_l(\boldsymbol \theta_{l})$.
   \STATE $\mathbf{B}_l=\mathbf{B}_{l+1}\mathbf{H}_{l}(\boldsymbol \theta_{l})$.
  \ENDFOR
  \STATE Construct $\mathbf{J}(\boldsymbol \theta)=\begin{bmatrix}
\mathbf{J}_1(\boldsymbol \theta)&\cdots&\mathbf{J}_L(\boldsymbol \theta)
\end{bmatrix}$
 \ENDIF
\STATE //PARAMETERS UPDATE
  \STATE Update $\boldsymbol \theta$.
    \ENDWHILE
    \STATE Compute the PhyCOM output $\widehat{\mathbf{s}}$.
 \RETURN $\widehat{\mathbf{s}}$, $\boldsymbol \theta$
 \end{algorithmic}\label{algo1}
 \end{algorithm}
 
   \begin{algorithm}[!t]
 \caption{PhyCOM Semi-Supervised Algorithm}
 \begin{algorithmic}[1]
 \renewcommand{\algorithmicrequire}{\textbf{Input:}}
 \renewcommand{\algorithmicensure}{\textbf{Output:}}
 \REQUIRE training data $\{\tilde{\mathbf{y}}_0,\tilde{\mathbf{x}}_{0P}\}$, allocation matrix $\mathbf{P}$, feedback. 
   \STATE  //SUPERVISED LEARNING
\STATE Apply the Algorithm 1 using the training data $\{\tilde{\mathbf{y}}_0,\tilde{\mathbf{x}}_{0P}\}$ and the allocation matrix $\mathbf{P}$.
 \STATE $\mathbf{P}\leftarrow \mathbf{I}_N$
\STATE Compute the PhyCOM output $\widehat{\mathbf{s}}$ and compute the corresponding symbols $\tilde{\mathbf{x}}_{0}=[\Re e^T(\widehat{\mathbf{s}}),\Im m^T(\widehat{\mathbf{s}})]^T$.
 \IF{feedback=True} 
  \STATE  //SELF TRAINING 
\STATE Apply the Algorithm 1 using the training data $\{\tilde{\mathbf{y}}_0,\tilde{\mathbf{x}}_{0}\}$ and the allocation matrix $\mathbf{I}_N$.
 \ENDIF
 \RETURN $\widehat{\mathbf{s}}$, $\boldsymbol \theta$
 \end{algorithmic}\label{algo2}
 \end{algorithm}

\subsubsection{Semi-Supervised Learning}
\label{sub_sec_imp}

Like other deep neural networks, the proposed architecture may suffer from overfitting for a small number of training symbols. To reduce overfitting, a simple solution is to exploit the alphabet $\mathcal{M}^N$ of the transmitted symbols.

To exploit the alphabet, we propose to use a self-labeled semi-supervised training approach \cite{TRI15}. During the supervised stage, the network is first trained using the knowledge of the $N_P\ll N$ pilot symbols. During the self training stage, the network is then retrained with the $N$ detected symbols~$\widehat{\mathbf{s}}$. The proposed semi-supervised algorithm is summarized in Algorithm~\ref{algo2}.

\subsubsection{Computational Complexity}

In this section, we report on the computational complexity of the training stage. In the following derivations, we assume that the communication chain is composed of $L$ layers. Because of the use of the LM optimization method, it is not possible to predict the number of iterations in Algorithm~\ref{algo1}. For this reason, we only provide some approximation results for one single iteration. 

In Algorithm~\ref{algo1}, the most computationally demanding tasks correspond to the matrix operations in lines 5, 14 and~15.
\begin{itemize}
\item In the forward pass, the computation of $\mathbf{H}_l(\boldsymbol \theta)$ requires a matrix inversion for each non-isomorphic layer. As the computational complexity of each inversion is $O((2N)^3)$, the total computational complexity of the forward pass is approximatively $O(8L_1N^3)$, where $L_1\le L$ corresponds to the number of non-isomorphic layers.
\item In the backward pass, the computation of $\mathbf{J}_l(\boldsymbol \theta_{l})$ has complexity $O((2N_p)(2N)(K_l))$ and the computation of $\mathbf{B}_l$ has complexity $O((2N_p)((2N)^2))$. When the total number of trainable parameters is significantly smaller than the number of samples (i.e. $K\ll N$), the computational complexity of the matrix product $\mathbf{J}_l(\boldsymbol \theta_{l})=\mathbf{B}_{l+1}\mathbf{L}_l(\boldsymbol \theta_{l})$ can be neglected. Therefore, the computational complexity of each backward pass is approximatively $O(8L N_p N^2)$.
\end{itemize}
The approximate computational complexity per iteration for the supervised and self training stages is presented in Table~\ref{summary_complexity}. It is important to note that the number of non-isomorphic layers, $L_1$,  has a significant impact on the computational complexity.

\section{SISO Communication Layers}
\label{commlayer}

This section presents a non-exhaustive list of commonly encountered layers in SISO communications. For each considered layer, it provides the mathematical expressions of the transfer matrix, compensation matrix, and local Jacobian. In the following, the layer index $l$ is dropped and the number of free parameters of each layer is denoted $K$ for the sake of simplicity.

\begin{table}[!t]
\centering
\begin{tabular}{|c||cc|}
\hline
& Forward Pass & Backward Pass \\
\hline
Supervised stage & $O(8L_1N^3)$ & $O(8L N_p N^2)$\\
Self training stage & $O(8L_1N^3)$& $O(8L N^3)$\\
\hline
\end{tabular}
\vskip 0.5em
\caption{Approximate Computational Complexity per Iteration ($L_1$ corresponds to the number of non-isomorphic layers).}\label{summary_complexity}
\end{table}

\subsection{IQ Imbalance Layer}

\subsubsection{Impairment Model}
IQ imbalance is usually modeled as \cite{VAL01,TAR05}
\begin{align}
y_l[n] = \alpha_1 y_{l-1}[n]+ \alpha_2 y_{l-1}^*[n],
\end{align}
where $n=0,1\cdots,N-1$ and $(\alpha_1,\alpha_2) \in \mathbb{C}^2$ are the IQ imbalance parameters. IQ imbalance layers depends on $4$ real-valued nuisance parameters, which are given by
\begin{align}
\boldsymbol \alpha =
\begin{bmatrix}
\alpha_1\\
\alpha_2\\
\alpha_3\\
\alpha_4
\end{bmatrix}=
\begin{bmatrix}
\Re e(\alpha_1+\alpha)\\
\Im m(-\alpha_1+\alpha)\\
\Im m(\alpha_1+\alpha)\\
\Re e(\alpha_1-\alpha)
\end{bmatrix}.
\end{align}
The transfer matrix of the IQ imbalance layer is given by
\begin{align}
\mathbf{F}(\boldsymbol \alpha)=\begin{bmatrix}
\alpha_1& \alpha_2\\
\alpha_3&\alpha_4
\end{bmatrix} \otimes \mathbf{I}_N\label{eqiq}.
\end{align}

\subsubsection{Compensation Model}
The IQ imbalance layer belongs to the class of isomorphic layers since $
\mathbf{F}(\boldsymbol \theta)\mathbf{F}(\boldsymbol \alpha)=\mathbf{I}_{2N}$ with
$\boldsymbol \theta=\mathbf{g}(\boldsymbol \alpha)=\frac{1}{\alpha_1\alpha_4-\alpha_2\alpha_3}\left[\alpha_4,-\alpha_2,-\alpha_3,\alpha_1\right]^T$. Using the Isomorphic trick, the compensation matrix is given by
\begin{align}
\mathbf{H}(\boldsymbol \theta)=\mathbf{F}(\boldsymbol \theta).
\end{align}
Using this simplification, the local Jacobian can also be expressed as 
\begin{align}
\mathbf{L}(\boldsymbol \theta)&=
\mathbf{I}_2 \otimes
\begin{bmatrix}\Re e(\mathbf{y}_{l-1})&\Im m(\mathbf{y}_{l-1})\end{bmatrix}.
\end{align}

\subsection{Carrier Frequency Offset Layer}

\subsubsection{Impairment Model}
The effect of a residual Carrier Frequency Offset (CFO) is usually modeled as \cite{YAO05}
\begin{align}
y_l[n] =y_{l-1}[n]e^{j\omega n},
\end{align}
where $\omega$ corresponds to the normalized residual carrier offset (in rad/samples). The CFO layer only depends on one layer parameter, which is given by $\boldsymbol \alpha=\omega$. Furthermore, this layer belongs to the class of strictly linear layers. Mathematically, the transfer matrix of the CFO layer can be expressed by 
$$
\mathbf{F}(\boldsymbol \alpha)=\underline{\mathbf{M}}(\boldsymbol \alpha),
$$
where
\begin{align}
\mathbf{M}(\boldsymbol\alpha)= 
\begin{bmatrix}
1 & 0 &\cdots & 0 \\
0 & e^{j \omega} &  &\vdots\\
 \vdots& & \ddots& 0\\
0&\cdots&0&e^{j \omega(N-1)}
\end{bmatrix}.
\end{align}
\subsubsection{Compensation Model}
The CFO layer also belongs to the class of isomorphic layers since $
\mathbf{F}(\boldsymbol \theta)\mathbf{F}(\boldsymbol\alpha)=\mathbf{I}_{2N}$ with $ \boldsymbol\theta=\mathbf{g}(\boldsymbol\alpha)=-\boldsymbol\alpha$. Using the isomorphic trick, the compensation matrix is then given by
\begin{align}
\mathbf{H}(\boldsymbol \theta)=\mathbf{F}(\boldsymbol \theta).
\end{align}
Using the expression of the compensation matrix, the local Jacobian can be expressed as
$$
\mathbf{L}(\boldsymbol \theta) = \tilde{\mathbf{q}}(\boldsymbol \theta),
$$
where $\mathbf{q}(\boldsymbol \theta)$ is a $N\times 1$ column vector defined as
\begin{align}
\mathbf{q}(\boldsymbol \theta)= 
 \begin{bmatrix}
0 \\
je^{j \theta} \\
\vdots \\
j(N-1)e^{j \theta(N-1)}
\end{bmatrix}
\odot \mathbf{y}_{l-1}.
\end{align}

\subsection{FIR Channel Layer}

\subsubsection{Impairment Model}
Let us consider a Finite Impulse Response (FIR) channel with $D$ taps. The output of an FIR channel layer is given by 
\begin{align}
y_l[n] = \sum_{d=0}^{D-1} h_d y_{l-1}[n-d],
\end{align}
where the complex-valued vector $\mathbf{h}=[h_0,\cdots,h_{D-1}]^T$ contains the channel coefficients. The FIR layer depends on $K=2D$ real-valued unknown parameters 
$\boldsymbol \alpha=\tilde{\mathbf{h}}$. This layer belongs to the class of strictly linear layer. Mathematically, the transfer matrix is given by
$$
\mathbf{H}(\boldsymbol \alpha)=\underline{\mathbf{M}}(\boldsymbol \alpha),
$$
where $\mathbf{M}(\boldsymbol \alpha)$ is a complex-valued Toeplitz matrix which is defined as
\begin{align}
\mathbf{M}(\boldsymbol  \alpha)=\begin{bmatrix}
h_0 & 0 & \cdots & 0 & 0\\
h_1 & h_0 &  \ddots &  & \vdots\\
h_2 & h_1 &  \ddots &  \ddots & \vdots\\
\vdots &  &  & \ddots & 0\\
0 & \cdots &  \cdots& \cdots & h_0\\
\end{bmatrix}.
\end{align}

\subsubsection{Compensation Model}
As the FIR channel layer does not satisfy the isomorphic equality, the compensation matrix must be computed from a matrix inversion as follows
\begin{align}
\mathbf{H}(\boldsymbol \theta)=\mathbf{F}^{-1}(\boldsymbol \theta).
\end{align}
As $\frac{d\mathbf{M}^{-1}(\boldsymbol  \theta)}{d\theta_k}=-\mathbf{M}^{-1}(\boldsymbol  \theta)\frac{d\mathbf{M}(\boldsymbol  \theta)}{d\theta_k}\mathbf{M}^{-1}(\boldsymbol  \theta)$, it can be checked that the local Jacobian is given by
\begin{align*}
\mathbf{L}(\boldsymbol \theta) = \begin{bmatrix}
\Re e(\mathbf{Q}(\boldsymbol \theta))\\
\Im m(\mathbf{Q}(\boldsymbol \theta))
\end{bmatrix},
\end{align*}
where $\mathbf{Q}(\boldsymbol \theta)=-\mathbf{M}^{-1}(\boldsymbol \theta)\left(\begin{bmatrix}
1&j \end{bmatrix}\otimes \mathbf{Y}_l\right)$ and
\begin{align}
\mathbf{Y}_l =\begin{bmatrix}
y_{l}[0] & 0 & \cdots &0 \\
y_{l}[1]&\ddots& \ddots&\vdots \\
\vdots && \ddots& 0\\
\vdots&  && y_{l}[0] \\
\vdots&  & & \vdots \\
y_{l}[N-1]&  \cdots& \cdots& y_{l}[N-1-D] \\
\end{bmatrix}.
\end{align}

\subsection{Quasi-Static Phase Noise}

\subsubsection{Impairment Model}
The effect of phase noise is usually modeled as \cite{KHA13,GHO17}
\begin{align}
y_{l}[n]=e^{j\varphi[n]}y_{l-1}[n],
\end{align}
where $\varphi[n]$ corresponds to the random phase. To limit the number of nuisance parameters, it is common to approximate phase noise by a Quasi-Static (QS) phase noise where $\varphi[n]$ is constant during $N/K$ samples. Under this assumption, the layer parameters are given by
$\boldsymbol \alpha = \begin{bmatrix}\varphi_1&\cdots&\varphi_{K}\end{bmatrix}^T$. 

The QS phase noise layer belongs to the class of strictly linear layers. Mathematically, the transfer matrix of the QS phase noise layer is given by 
\begin{align*}
\mathbf{F}(\boldsymbol \alpha)=\underline{\mathbf{M}}(\boldsymbol \alpha),
\end{align*}
where 
\begin{align}
\mathbf{M}(\boldsymbol \alpha)= 
\begin{bmatrix}
e^{j \varphi_1} & 0 &\cdots & 0 \\
0 & e^{j \varphi_2}&  &\vdots\\
\vdots & & \ddots& 0\\
0&\cdots&0&e^{j \varphi_K}
\end{bmatrix}\otimes \mathbf{I}_{N/K}.
\end{align}

\subsubsection{Compensation Model}
As $\mathbf{F}(\boldsymbol  \theta)\mathbf{F}(\boldsymbol  \alpha)$ with $\boldsymbol \theta=-\boldsymbol \alpha$, the Quasi-Static phase noise layer also belongs to the class of isomorphic layers. The compensation matrix is then given by
\begin{align}
\mathbf{H}(\boldsymbol \theta)=\mathbf{F}(\boldsymbol \theta).
\end{align}
Using this simplification, the local Jacobian can be expressed as
\begin{align*}
\mathbf{L}(\boldsymbol \theta) = \begin{bmatrix}
\Re e(\mathbf{Q}(\boldsymbol \theta))\\
\Im m(\mathbf{Q}(\boldsymbol \theta))
\end{bmatrix},
\end{align*}
where $\mathbf{Q}(\boldsymbol \theta)=
\begin{bmatrix}
\mathbf{q}_1(\boldsymbol \theta)\odot \mathbf{y}_{l-1} & \cdots & \mathbf{q}_K(\boldsymbol \theta)\odot \mathbf{y}_{l-1}
\end{bmatrix}$ with
\begin{align}
\mathbf{q}_k(\boldsymbol \theta)
&= \begin{bmatrix}
je^{j \varphi_1}\delta[k-1]\\
je^{j \varphi_2}\delta[k-2]\\
\vdots \\
je^{j \varphi_K}\delta[k-K]
\end{bmatrix}\otimes \mathbf{1}_{N/K},\end{align}
and $\delta[n]$ corresponds to the unit sample function that is equal to $1$ at $n=0$ and is zero elsewhere.

\section{Simulation Results}
\label{simu}

This section highlights the proposed algorithm's performance for symbol detection under unknown physical impairments. The proposed algorithm has been implemented using Python with the scientific \texttt{Numpy} / \texttt{Scipy} libraries~\cite{HAR20,VIR20}. The source code is available on Github at \url{https://github.com/vincentchoqueuse/PhyCOM}. All the simulations have been run on the cloud using AWS EC2 \texttt{t2.large} instances. 

In the following simulations, we have considered two different physical channel models. For these two models, the input vector $\mathbf{s}$ is composed of $N=500$ symbols generated from a 16-QAM constellation. Each input vector contains $N_p\ll N$ pilot samples, $\mathbf{s}_{0P}$, that are known at the receiver. The input vector is partitioned into two sets. The $N_p$ pilots belong to the training set, and the remaining $N-N_p$ data samples belong to the testing set.

The following simulations focus on the Mean-Squares Error (MSE) and Symbol Error Rate (SER) metrics to evaluate the proposed technique's performance. These metrics are evaluated for the training and testing sets. Let us denote by $\mathbf{P}$ and $\mathbf{P}^{\perp}$ the allocation matrices that extract the pilot samples and data samples, respectively. For the training set, the MSE and SER are given by
\begin{align}
\text{MSE}&=\frac{1}{N_p}E[\| (\mathbf{I}_2\otimes \mathbf{P}) \left(\tilde{\mathbf{x}}_{0} -\tilde{\mathbf{y}}_{L}\right)\|^2_2],\\
\text{SER}&=\frac{1}{N_p}E[\|  \mathbf{P}\left(\mathbf{s} -\widehat{\mathbf{s}} \right)\|_0].
\end{align}
where $\|.\|_0$ corresponds to the $\mathcal{L}_0$-norm. In each simulation, the MSE and SER are estimated using $100$ Monte Carlo trials. 

For the testing set, the MSE and SER are obtained by replacing $N_p$ with $N-N_p$ and $\mathbf{P} $ with $\mathbf{P}^{\perp}$ in the above expressions. Even if the MSE and SER for the training set give some information about the training stage, it should be emphasized that the main objective is to minimize these metrics for the testing set. For this reason, most of the following figures report on the MSE and SER for the testing set.

In each experiment, the last layer of the physical channel is a Gaussian noise layer with distribution $\tilde{\mathbf{b}}\sim \mathcal{N}(0,\frac{\sigma^2}{2}\mathbf{I}_{2N})$. For comparison purposes, the performance of the trained PhyCOM is compared with the performance of the clairvoyant PhyCOM with perfect knowledge of the channel model parameters $\boldsymbol \alpha$. For the clairvoyant PhyCOM, it can be checked that the theoretical MSE is given by
\begin{align}
\textrm{MSE}_{theo}=\frac{\sigma^2}{2N}\textrm{Tr}\left[\mathbf{F}_{tot}^{-1}(\boldsymbol \alpha)\mathbf{F}_{tot}^{-T}(\boldsymbol \alpha)\right]
\end{align}
where $\textrm{Tr}[.]$ corresponds to the matrix trace, and $\sigma^2$ is the noise variance.

\subsection{Simple Physical Communication Model}

\begin{figure*}[!t]
\centering
\begin{tikzpicture}[node distance=7em,scale=0.75, every node/.style={scale=0.75}]
\node (start) {$\mathbf{s}$};
\node [rectangle,draw,minimum width=4em,minimum height=4em,text width=4em,text centered,right of=start,node distance=5em](iq1){$\Re e$ / $\Im m$ \\ Splitting};
\node [rectangle,draw,minimum width=4em,minimum height=4em,text width=4em,text centered,right of=iq1](fir){FIR\\Channel};
\node [rectangle,draw,minimum width=4em,minimum height=4em,text width=4em,text centered,right of=fir](cfo2){CFO};
\node [rectangle,draw,minimum width=4em,minimum height=4em,text width=4em,text centered,right of=cfo2](iq2){IQ Rx};
\node [rectangle,draw,minimum width=4em,minimum height=4em,text width=4em,text centered,right of=iq2](noise){Noise};
\node [rectangle,draw,minimum width=4em,minimum height=4em,text width=4em,text centered,below of=iq2](iq2r){IQ Rx};
\node [rectangle,draw,minimum width=4em,minimum height=4em,text width=4em,text centered,below of=cfo2](cfo2r){CFO};
\node [rectangle,draw,minimum width=4em,minimum height=4em,text width=4em,text centered,below of=fir](firr){FIR\\Channel};
\node [rectangle,draw,minimum width=4em,minimum height=4em,text width=4em,text centered,below of=iq1](iq1r){Non-Linear Detection};
\node [left of=iq1r,node distance=5em] (output) {$\widehat{\mathbf{s}}$};
\draw[dashed] ($ (iq1) - (1.2,-1) $)  rectangle ($ (noise) + (1.2,-1) $)  node[midway,yshift=3.5em]{Physical Channel Model};
\draw[dashed] ($ (iq1r) - (1.2,-1) $)  rectangle ($ (iq2r) + (1.2,-1) $)  node[midway,yshift=-3.5em]{PhyCOM};
\draw[->,>=latex] (start) -- (iq1);
\draw[->,>=latex] (iq1) -- (fir);
\draw[->,>=latex] (fir) -- (cfo2);
\draw[->,>=latex] (cfo2) -- (iq2);
\draw[->,>=latex] (iq2) -- (noise);
\draw[->,>=latex] (noise)--++(1.5,0) |- (iq2r) node[above,xshift=5em] {$\tilde{\mathbf{y}}_0$};
\draw[->,>=latex] (iq2r) -- (cfo2r);
\draw[->,>=latex] (cfo2r) -- (firr);
\draw[->,>=latex] (firr) -- (iq1r);
\draw[->,>=latex] (iq1r) -- (output);
\end{tikzpicture}
\caption{Simple Physical Communication Model. The physical layer is composed of a FIR channel composed of $8$ taps ($\mathbf{h}=[0.9+0.1j,0.3+0.3j,0.1+0.05j,0.02+0.1j,0.1-0.05j,0.02-0.1j,0.1+0.03j,0.04-0.012j]^T$), a receiver with CFO ($\boldsymbol \theta=0.005$) and IQ impairment ($\boldsymbol \theta=[1.8,0.1,0.13,0.8]^T$). The received signal is corrupted by a circular white Gaussian noise $\mathcal{N}_{\mathcal{C}}(0,\sigma^2)$.}\label{wirelesschannel}
\end{figure*}
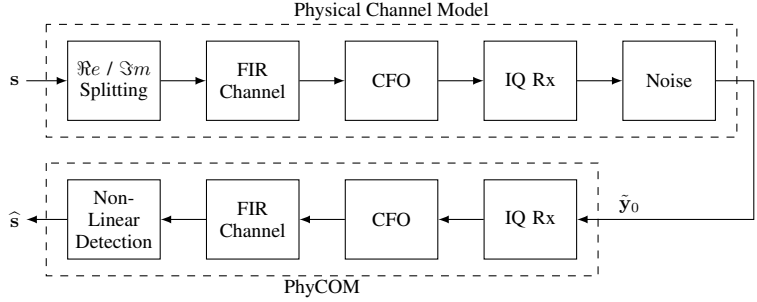

In the first experiment, we consider a simple communication model composed of a FIR channel composed of $8$ taps, a Carrier Frequency Offset, an IQ imbalance, and a Gaussian noise layer. The considered communication chain and the associated PhyCOM network are described in Fig.\ref{wirelesschannel}. The PhyCOM network depends on $K=13$ static parameters. In this context, the $N_p$ pilot samples are allocated using a preamble-based strategy (see Fig.\ref{fig_allocate}).

\subsubsection{One-Shot Analysis}

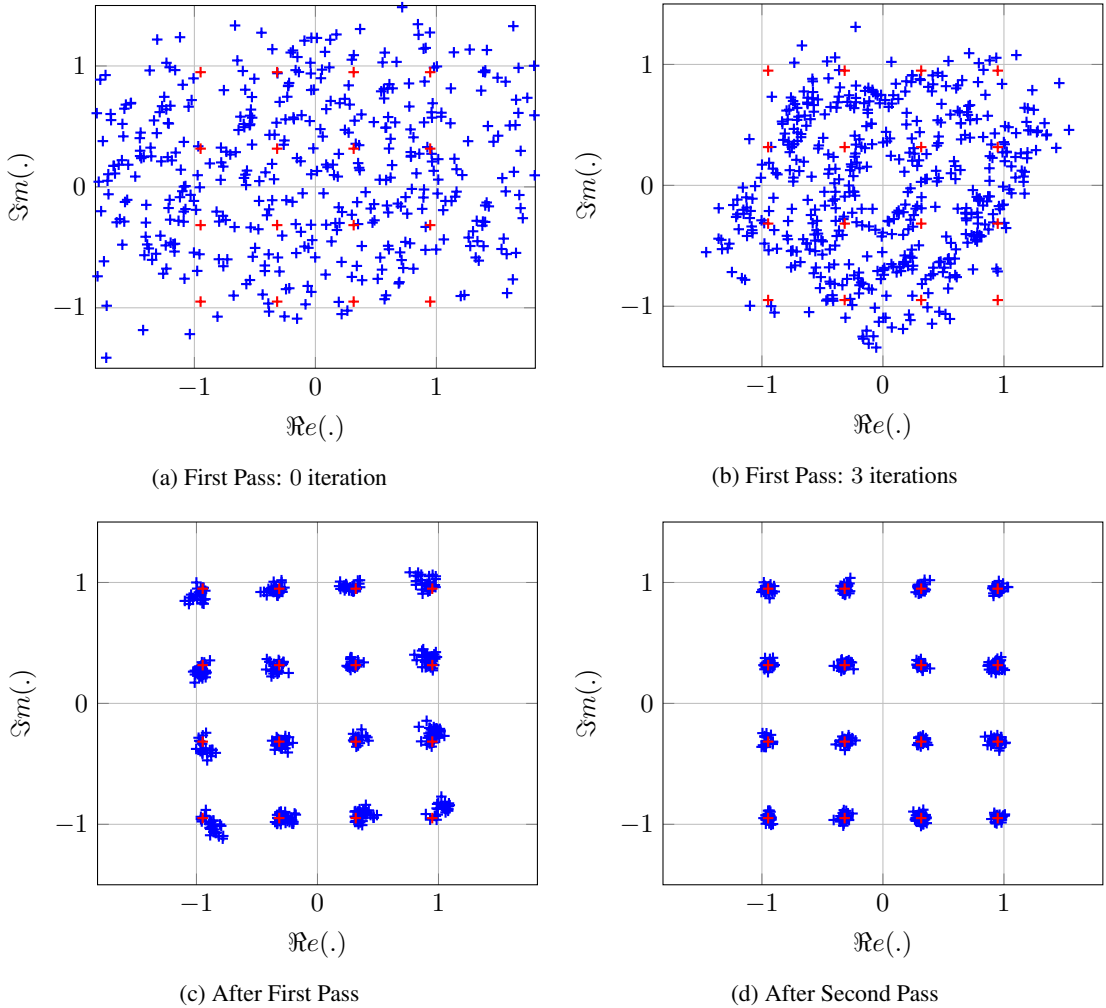
\begin{figure}[t]
\begin{subfigure}{.45\linewidth}
\centering
\begin{tikzpicture}
    \begin{axis}[complexplot]
         \addplot+[thick,only marks,mark=+] table[x index=1, y index=2, col sep=comma] {./csv/one_shot/constellation_before.csv}; 
             \addplot[thick,only marks,red,mark=+] coordinates {(-0.9486833,-0.9486833)(-0.9486833,-0.31622777)(-0.9486833,0.9486833)
(-0.9486833,0.31622777)(-0.31622777,-0.9486833)(-0.31622777,-0.31622777)
(-0.31622777,0.9486833)(-0.31622777,0.31622777)(0.9486833,-0.9486833)
 (0.9486833,-0.31622777)(0.9486833,0.9486833)(0.9486833,0.31622777)
 (0.31622777,-0.9486833)(0.31622777,-0.31622777)(0.31622777,0.9486833)
 (0.31622777,0.31622777)}; 
    \end{axis}
\end{tikzpicture}
\caption{First Pass: $0$ iteration}
\end{subfigure}
\begin{subfigure}{.45\linewidth}
\centering
\begin{tikzpicture}
    \begin{axis}[complexplot]
         \addplot+[thick,only marks,mark=+] table[x index=1, y index=2, col sep=comma]  {./csv/one_shot/constellation_after_training_step1_3.csv}; 
             \addplot[thick,only marks,red,mark=+] coordinates {(-0.9486833,-0.9486833)(-0.9486833,-0.31622777)(-0.9486833,0.9486833)
(-0.9486833,0.31622777)(-0.31622777,-0.9486833)(-0.31622777,-0.31622777)
(-0.31622777,0.9486833)(-0.31622777,0.31622777)(0.9486833,-0.9486833)
 (0.9486833,-0.31622777)(0.9486833,0.9486833)(0.9486833,0.31622777)
 (0.31622777,-0.9486833)(0.31622777,-0.31622777)(0.31622777,0.9486833)
 (0.31622777,0.31622777)}; 
    \end{axis}
\end{tikzpicture}
\caption{First Pass: $3$ iterations}
\end{subfigure}
\vskip 1em
\begin{subfigure}[t]{.45\linewidth}
\centering
\begin{tikzpicture}
    \begin{axis}[complexplot]
         \addplot+[thick,only marks,mark=+] table[x index=1, y index=2, col sep=comma]  {./csv/one_shot/constellation_after_training_step1.csv}; 
             \addplot[thick,only marks,red,mark=+] coordinates {(-0.9486833,-0.9486833)(-0.9486833,-0.31622777)(-0.9486833,0.9486833)
(-0.9486833,0.31622777)(-0.31622777,-0.9486833)(-0.31622777,-0.31622777)
(-0.31622777,0.9486833)(-0.31622777,0.31622777)(0.9486833,-0.9486833)
 (0.9486833,-0.31622777)(0.9486833,0.9486833)(0.9486833,0.31622777)
 (0.31622777,-0.9486833)(0.31622777,-0.31622777)(0.31622777,0.9486833)
 (0.31622777,0.31622777)}; 
    \end{axis}
\end{tikzpicture}
\caption{After First Pass}
\end{subfigure}
\begin{subfigure}[t]{.45\linewidth}
\centering
\begin{tikzpicture}
    \begin{axis}[complexplot]
         \addplot+[thick,only marks,mark=+] table[x index=1, y index=2, col sep=comma]  {./csv/one_shot/constellation_after_training_step2.csv}; 
             \addplot[thick,only marks,red,mark=+] coordinates {(-0.9486833,-0.9486833)(-0.9486833,-0.31622777)(-0.9486833,0.9486833)
(-0.9486833,0.31622777)(-0.31622777,-0.9486833)(-0.31622777,-0.31622777)
(-0.31622777,0.9486833)(-0.31622777,0.31622777)(0.9486833,-0.9486833)
 (0.9486833,-0.31622777)(0.9486833,0.9486833)(0.9486833,0.31622777)
 (0.31622777,-0.9486833)(0.31622777,-0.31622777)(0.31622777,0.9486833)
 (0.31622777,0.31622777)}; 
    \end{axis}
\end{tikzpicture}
\caption{After Second Pass}
\end{subfigure}
\caption{Constellation of $\mathbf{P}^{\dagger}\mathbf{y}_{L}$ ($N_P=50$ symbols, SNR=$30$dB, testing set).}\label{dataconst}
\end{figure}

\begin{figure}[!t]
\begin{subfigure}[t]{1.\linewidth}
\centering
\begin{tikzpicture}
    \begin{semilogyaxis}[height=0.35\linewidth,width=0.75\linewidth,xlabel near ticks,ylabel near ticks,grid=both,ymax=2,ymin=0.0008,xmin=0,ylabel=MSE,xlabel=Number of iterations,xmax=19]
    \addplot+[thick,black,mark=none] coordinates {(0,0.00165)(19,0.00165)};
    \addplot+[thick,blue,dashed,mark=none] table[x index=0, y index=1, col sep=comma] {./csv/one_shot/mse_first_pass.csv}; 
     \addplot+[thick,blue,mark=none] table[x index=0, y index=2, col sep=comma] {./csv/one_shot/mse_first_pass.csv}; 
       \draw[dashed,black,thick] (axis cs:26,0.0003) -- (axis cs:26,10);
       \draw[<->,>=latex,red,thick] (axis cs:18,0.0010) -- node[midway](node1){} (axis cs:18,0.008);
       \draw[->,>=latex,red,thick] (axis cs:10,0.012) node[above]{Generalization error}-- (node1);
     \legend{Clairvoyant,Training Data, Testing Data}
      \end{semilogyaxis}
\end{tikzpicture}
\caption{Step 1: Supervised stage.}\label{mse1fig00}
\end{subfigure}
\begin{subfigure}[t]{1.\linewidth}
\centering
\begin{tikzpicture}
    \begin{semilogyaxis}[height=0.35\linewidth,width=0.75\linewidth,xlabel near ticks,ylabel near ticks,grid=both,ymax=2,ymin=0.0008,xmin=0,ylabel=MSE,xlabel=Number of iterations,xmax=14]
     \addplot+[thick,black,mark=none] coordinates {(0,0.00165)(14,0.00165)}; 
    \addplot+[thick,blue,dashed,mark=none] table[x index=0, y index=1, col sep=comma] {./csv/one_shot/mse_second_pass.csv}; 
     \addplot+[thick,blue,mark=none] table[x index=0, y index=2, col sep=comma] {./csv/one_shot/mse_second_pass.csv}; 
       \draw[dashed,black,thick] (axis cs:26,0.0005) -- (axis cs:26,10);
     \legend{Clairvoyant,Training Data, Testing Data}
      \end{semilogyaxis}
\end{tikzpicture}
\caption{Step 2: Self training stage.}\label{mse1fig01}
\end{subfigure}
\caption {Evolution of the MSE during learning ($N_P=50$ symbols, SNR=$30$dB).\label{mse1fig0tot}}
\end{figure}
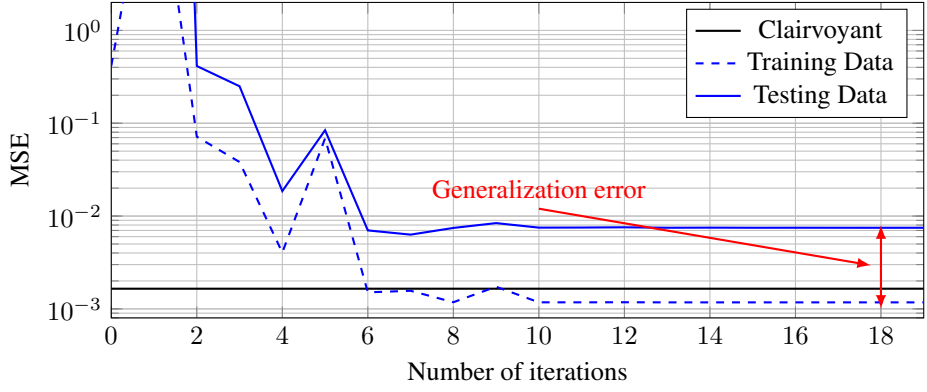
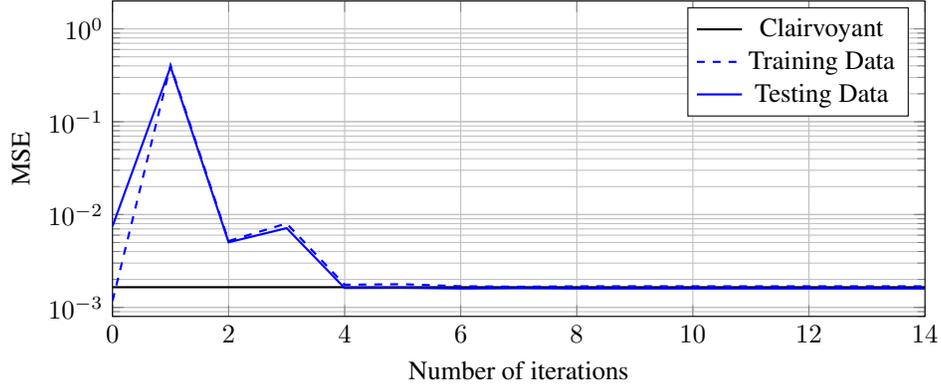

\pgfplotscreateplotcyclelist{mark list}{%
    every mark/.append style={solid,fill=black},mark=o,only marks,thick\\%
    every mark/.append style={solid,fill=black},mark=none,thick,black\\%
    every mark/.append style={solid,fill=blue},mark=none,thick,blue,solid\\%
    every mark/.append style={solid,fill=blue},mark=none,thick,blue,dashed\\%
    every mark/.append style={solid,fill=blue},mark=none,thick,blue,dashdotted\\%
  every mark/.append style={solid,fill=blue},mark=diamond,thick,blue\\%
    every mark/.append style={solid,fill=red},mark=none,thick,red,dashed\\%
    every mark/.append style={solid,fill=red},mark=+,thick,red,solid\\%
}

\begin{figure}[!t]
\centering
\begin{tikzpicture}
    \begin{semilogyaxis}[width=0.5\textwidth, xlabel near ticks,ylabel near ticks,grid=both,ylabel=MSE,xlabel=$N_p$,xmin=14,ymax=1,xmax=80, ymin=0.0005,cycle list name=mark list]
    	\addplot table[x index=0, y=theo, col sep=comma] {./csv/simulation2/mse_test.csv}; 				
    	\addplot table[x index=0, y=clairvoyant, col sep=comma] {./csv/simulation2/mse_test.csv}; 				
      	\addplot table[x index=0, y=old1, col sep=comma] {./csv/simulation2/mse_test.csv}; 				
 	\addplot table[x index=0, y=old2, col sep=comma] {./csv/simulation2/mse_test.csv}; 		
     	\addplot table[x index=0, y=old3, col sep=comma] {./csv/simulation2/mse_test.csv}; 			
	\addplot table[x index=0, y=old4, col sep=comma] {./csv/simulation2/mse_test.csv}; 
	\addplot table[x index=0, y=phycom1, col sep=comma] {./csv/simulation2/mse_test.csv}; 
	\addplot table[x index=0, y=phycom2, col sep=comma] {./csv/simulation2/mse_test.csv}; 
    \legend{Theoretical, Clairvoyant, DSP1, DSP2, DSP3, DSP4,  PhyCOM1, PhyCOM2}
    \end{semilogyaxis}
\end{tikzpicture}
\caption{Evolution of the training and testing MSE versus number of pilots (SNR$=30$ dB).}\label{full1fig}
\end{figure}
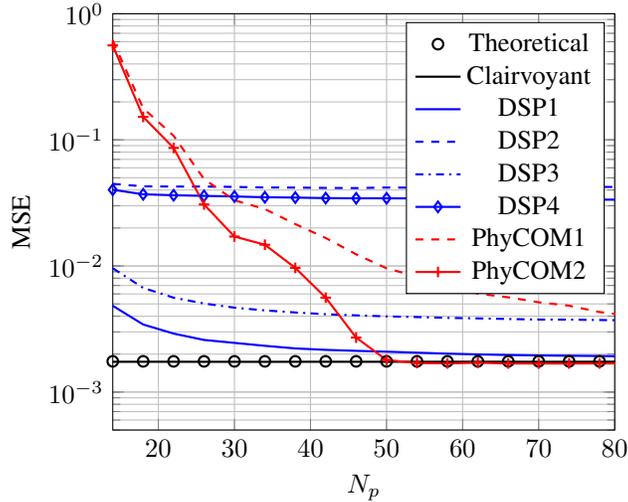

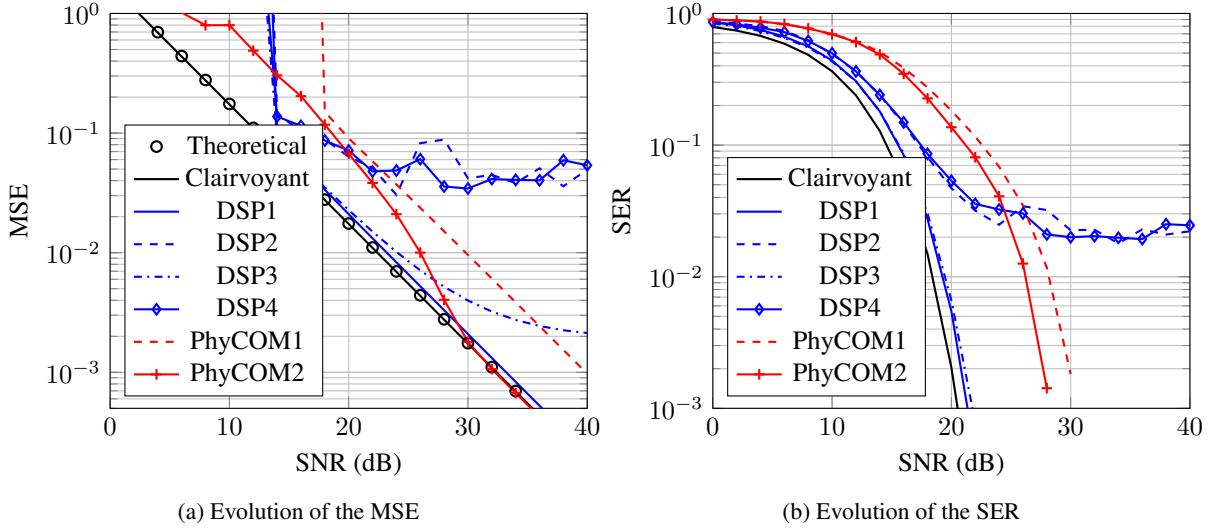
\begin{figure}[!t]
\centering
\begin{subfigure}{.48\textwidth} 
\centering
\begin{tikzpicture}
    \begin{semilogyaxis}[width=1.0\textwidth,xlabel near ticks,ylabel near ticks,grid=both,ylabel=MSE,xlabel=SNR (dB),xmin=0,ymax=1,xmax=40,legend pos=south west, ymin=0.0005,cycle list name=mark list]
    	\addplot table[x index=0, y=theo, col sep=comma] {./csv/simulation3/mse_test.csv}; 				
    	\addplot table[x index=0, y=clairvoyant, col sep=comma] {./csv/simulation3/mse_test.csv}; 				
      	\addplot table[x index=0, y=old1, col sep=comma] {./csv/simulation3/mse_test.csv}; 				
 	\addplot table[x index=0, y=old2, col sep=comma] {./csv/simulation3/mse_test.csv}; 		
     	\addplot table[x index=0, y=old3, col sep=comma] {./csv/simulation3/mse_test.csv}; 			
	\addplot table[x index=0, y=old4, col sep=comma] {./csv/simulation3/mse_test.csv}; 
	\addplot table[x index=0, y=phycom1, col sep=comma] {./csv/simulation3/mse_test.csv}; 
	\addplot table[x index=0, y=phycom2, col sep=comma] {./csv/simulation3/mse_test.csv}; 
    \legend{Theoretical, Clairvoyant, DSP1, DSP2, DSP3, DSP4, PhyCOM1, PhyCOM2}
    \end{semilogyaxis}
\end{tikzpicture}
\caption{Evolution of the MSE}\label{mse1fig2}
\end{subfigure}
\begin{subfigure}{.48\textwidth}
\centering
\begin{tikzpicture}
    \begin{semilogyaxis}[width=1.0\textwidth,xlabel near ticks,ylabel near ticks,grid=both,ylabel=SER,xlabel=SNR (dB),xmin=0,ymax=1,xmax=40,ymin=0.001,legend pos=south west,, cycle list name=mark list,cycle list shift=1]
    	\addplot table[x index=0, y=clairvoyant, col sep=comma] {./csv/simulation3/ser_test.csv}; 				
      	\addplot  table[x index=0, y=old1, col sep=comma] {./csv/simulation3/ser_test.csv}; 				
 	\addplot table[x index=0, y=old2, col sep=comma] {./csv/simulation3/ser_test.csv}; 		
     	\addplot table[x index=0, y=old3, col sep=comma] {./csv/simulation3/ser_test.csv}; 			
	\addplot table[x index=0, y=old4, col sep=comma] {./csv/simulation3/ser_test.csv}; 
	\addplot table[x index=0, y=phycom1, col sep=comma] {./csv/simulation3/ser_test.csv}; 
	\addplot table[x index=0, y=phycom2, col sep=comma] {./csv/simulation3/ser_test.csv}; 		
    \legend{Clairvoyant, DSP1, DSP2, DSP3, DSP4, PhyCOM1, PhyCOM2}
    \end{semilogyaxis}
\end{tikzpicture}
\caption{Evolution of the SER}\label{ser1fig2}
\end{subfigure}
\caption{Evolution of the training and testing errors versus SNR ($N_P=50$ symbols).}\label{full2fig}
\end{figure}

In the first experiment, we consider an input vector of length $N=500$ composed of $N_P=50$ training samples and a channel model with a Signal to Noise Ratio (SNR) equal to $30$dB. In the PhyCOM network, the IQ and CFO layers are initialized with $\boldsymbol\theta_1=\mathbf{0}$ and $\boldsymbol\theta_2=\mathbf{0}$, respectively, and the FIR layer is initialized with $\mathbf{h}=[1,0,0,0,0,0,0,0]$.

Fig.~\ref{dataconst} presents the constellation of the estimated data, $\mathbf{P}^{\dagger}\mathbf{y}_{L}$, after the supervised and self training stages. After the supervised step, the constellation seems to be distorted by a residual phase rotation. After the self training stage, we observe that this rotation is corrected and that the constellation of the estimated data perfectly matches the 16-QAM constellation.

To highlight the contribution of the self training stage, Fig.~\ref{mse1fig0tot} reports on the evolution of the Mean-Squares Error (MSE) during the learning stage. The dashed and solid lines depict the MSE for the training and testing sets, respectively. Fig.~\ref{mse1fig00} shows that the proposed algorithm converges in $19$ iterations during the supervised step. This figure also shows that the testing MSE is higher than the MSE of the clairvoyant PhyCOM and that there is a significant gap between the training and testing MSEs. This gap is often called the generalization error and indicates that the supervised step tends to overfit the training set. Fig.~\ref{mse1fig01} presents the evolution of the MSE during the self training stage. This figure shows that the testing MSE significantly decreases between the supervised and self training stages. After $14$ iterations, the training MSE, the testing MSE and the MSE of the clairvoyant PhyCOM are nearly identical. This behavior highlights the fact the semi-supervised learning makes the PhyCOM network less prone to overfitting.

\subsubsection{Comparison with Conventional Techniques}

\begin{table}[!t]
\centering
\begin{tabular}{|c|ccc|c|}
\hline
Technique & IQ & CFO & FIR & Self Training\\
\hline
Clairvoyant & Known & Known & Known & No \\
DSP1& Known & Known & Trained & No \\
DSP2& Known & Blind \cite{SEL09} & Trained \cite{KAY93} & No \\
DSP3& Blind \cite{VAL05}& Known & Trained \cite{KAY93}  & No\\
DSP4& Blind \cite{VAL05} & Blind \cite{SEL09}& Trained \cite{KAY93} & No \\
PhyCOM1 &  \multicolumn{3}{c|}{Jointly Trained } & No \\
PhyCOM2 &  \multicolumn{3}{c|}{Jointly Trained } & Yes \\
\hline
\end{tabular}
\vskip 0.5em
\caption{List of Considered Algorithms.}\label{consalgo}
\end{table}

In the following simulations, we compare the performance of the PhyCOM algorithm with other techniques. Here, we choose to report on the performance of conventional Digital Signal Processing (DSP) techniques since more sophisticated deep learning approaches using non-parametric channel estimation require a larger training set to be usable. For example, a naive implementation of the MMSE channel estimator used in \cite{HE20} requires at least $(2N)^2=10^6$ real-valued pilots for the estimation of the $2N\times 2N$ accumulated transfer matrix. The considered DSP techniques for the compensation of the IQ imbalance, CFO, and FIR channel, are presented in Table~\ref{consalgo}. For blind IQ imbalance compensation, we have implemented a standard algorithm based on the diagonalization of the augmented second-order statistics~\cite{VAL05,FAT08}. For CFO estimation, we have implemented a blind method based on the maximization of the fourth-order statistics~\cite{SEL09}\footnote{For periodogram maximization, we have implemented a Newton optimization algorithm that is initialized with the true value of $\omega$.}. For FIR channel estimation and compensation, we have implemented a trained-based algorithm based on the least-squares technique (see Example 4.13 in \cite{KAY93}), that uses a polynomial division for deconvolution. Using these techniques, we have considered four conventional DSP compensators, which are described in Table~\ref{consalgo}. Note that DSP1, DSP2, and DSP3 techniques are partially clairvoyant approaches since they require the knowledge of some parameters. The DSP4 technique is the only DSP technique that does not require the knowledge of some layer parameters. In Table~\ref{consalgo}, PhyCOM1 and PhyCOM2 techniques correspond to the proposed PhyCOM approach using a supervised-only or a semi-supervised training strategy, respectively.

Fig.~\ref{full1fig} presents the evolution of the MSE and SER versus the number of pilot samples $N_p$. The black curve shows the performance of the clairvoyant PhyCOM. We observe that the PhyCOM2 is the only technique that achieves near-optimal performance. The DSP2 and DSP4 techniques do not achieve satisfactory performance. More precisely, the MSE is equal to $0.042$ and $0.033$ for the DSP2 and DSP4 techniques, respectively. By analyzing the performance of the other DSP techniques, we observe that the poor performance of the DSP2 and DSP4 techniques is due to the blind CFO estimator. The performance of the blind CFO estimator critically depends on the number of samples $N$. For example, additional experiments have shown that the MSEs of the DSP2 and DSP4 techniques decrease to $0.0075$ and $0.0083$ for $N=1000$ samples, and to $0.0047$ and $0.0050$ for $N=2000$ samples.

Table~\ref{contime} reports on the computation complexity of the DSP and PhyCOM techniques. We observe that the PhyCOM1 and PhyCOM2 techniques are more computationally demanding than the conventional DSP techniques. Table~\ref{table_time2} presents the training time associated with the supervised and self training stages. This table also reports on the average number of iterations and training time per iteration. For the supervised stage, we observe that the total time for the forward pass is significantly larger than for the backward pass because of the matrix inversion of the FIR channel layer. For the self training stage, we note that the total time for the forward and backward passes are broadly similar since the cost of the Jacobian propagation becomes more significant when $N_p=N$. It should be emphasized that several strategies can be employed to reduce the training time. First, for preamble-based allocation, it is possible to truncate the input vector $\tilde{\mathbf{x}}_0$ to $N_t$ samples ($N_p\le N_t<N$). For example, setting $N_t=N_p=80$ samples does not modify the MSE performance of the PhyCOM1 technique but decreases the average training time to $0.042$s. For the PhyCOM2 technique, truncating the number of samples to $N=100$ samples slightly increases the MSE to $0.00299$ (instead of $0.00168$) but reduces the average training time to $0.078$s (instead of $2.848$s). In practical applications, it is also possible to freeze some static non-isomorphic layers when their parameters have been previously estimated. 

\begin{table}[!t]
\centering
\begin{tabular}{|c|cccccc|}
\hline
& DSP1 & DSP2 & DSP3 & DSP4 & PhyCOM1 & PhyCOM2\\
\hline
Time (s) & $0.002$ & $0.003$ & $0.002$  & $0.003$ & $1.344$ & $2.848$\\
\hline
\end{tabular}
\vskip 0.5em
\caption{Average Training Time in Seconds ($N_P=80$ symbols, $N=500$ samples, SNR=$30$dB). }\label{contime}
\end{table}

\begin{table}[!t]
\centering
\begin{tabular}{|c|c||c|cc|}
\hline
Stage &Pass & Total Time & Av. Nb it.& Av Time per it.\\
\hline
\multirow{2}{*}{Supervised}&Forward & 1.142 (s) & 21.75 & 0.052 (s)\\
&Backward& 0.201 (s) & 13.14  & 0.015 (s)\\
\hline
\hline
\multirow{2}{*}{Self Training}&Forward & 0.827 (s) & 16.44 & 0.050 (s)\\
&Backward & 0.676 (s) & 8.78  & 0.076 (s)\\
\hline
\end{tabular}
\vskip 0.5em
\caption{Training Time ($L_1=1$ non-isomorphic layer, $N_p=80$ symbols, $N=500$ samples).}\label{table_time2}
\end{table}

Fig.~\ref{full2fig} reports on the MSE and SER performances versus Signal to Noise Ratio (SNR) for $N_p=50$ training symbols. For SNR$<18$ dB, Fig.~\ref{mse1fig2} shows that the proposed technique cannot achieve an acceptable MSE. For SNR$\ge 20$ dB, we observe that the testing MSE for the PhyCOM1 technique is parallel to those of the clairvoyant PhyCOM ($8$dB penalty). For the PhyCOM2 technique, the testing MSE approaches the one obtained with the clairvoyant PhyCOM with SNR$\ge 30$dB. Fig.~\ref{ser1fig2} also shows the same behavior for the testing SER. We also note that the partially clairvoyant DSP1 and DSP3 techniques give satisfactory results. However, these techniques assume at least a perfect knowledge of the CFO parameter. 

As compared to conventional DSP techniques, the results reported here have shown that the PhyCOM technique offers the best statistical performance both in terms of MSE and SER under the blind scenario, at the expense of increased complexity. Nevertheless, it should be noted that the DSP and PhyCOM approaches are not incompatible in practical applications. For example, it is possible to combine the best of both worlds by using first the PhyCOM approach for initial parameter estimation, followed by a clairvoyant DSP technique for low-complexity compensation.

\subsection{Performance With Time-Varying Parameters and Model Mismatch}

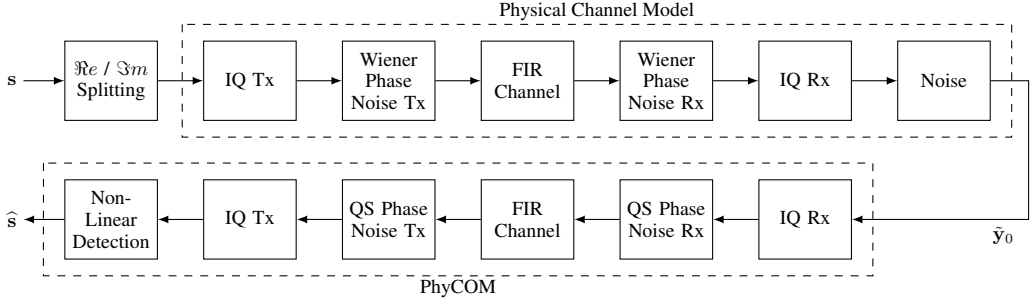
\begin{figure*}[!t]
\centering
\begin{tikzpicture}[node distance=7em,scale=0.75, every node/.style={scale=0.75}]. 
\node (start) {$\mathbf{s}$};
\node [rectangle,draw,minimum width=4em,minimum height=4em,text width=4em,text centered,right of=start,node distance=5em](mod){$\Re e$ / $\Im m$ \\ Splitting};
\node [rectangle,draw,minimum width=4em,minimum height=4em,text width=4em,text centered,right of=mod](iq1){IQ Tx};
\node [rectangle,draw,minimum width=4em,minimum height=4em,text width=4em,text centered,right of=iq1](pn1){Wiener Phase Noise Tx};
\node [rectangle,draw,minimum width=4em,minimum height=4em,text width=4em,text centered,right of=pn1](fir){FIR\\Channel};
\node [rectangle,draw,minimum width=4em,minimum height=4em,text width=4em,text centered,right of=fir](pn2){Wiener Phase Noise Rx};
\node [rectangle,draw,minimum width=4em,minimum height=4em,text width=4em,text centered,right of=pn2](iq2){IQ Rx};
\node [rectangle,draw,minimum width=4em,minimum height=4em,text width=4em,text centered,right of=iq2](noise){Noise};
\node [rectangle,draw,minimum width=4em,minimum height=4em,text width=4em,text centered,below of=iq2](iq2r){IQ Rx};
\node [rectangle,draw,minimum width=4em,minimum height=4em,text width=4em,text centered,below of=pn2](pn2r){QS Phase Noise Rx};
\node [rectangle,draw,minimum width=4em,minimum height=4em,text width=4em,text centered,below of=fir](firr){FIR\\Channel};
\node [rectangle,draw,minimum width=4em,minimum height=4em,text width=4em,text centered,below of=pn1](pn1r){QS Phase Noise Tx};
\node [rectangle,draw,minimum width=4em,minimum height=4em,text width=4em,text centered,below of=iq1](iq1r){IQ Tx};
\node [rectangle,draw,minimum width=4em,minimum height=4em,text width=4em,text centered,below of=mod](det){Non- Linear Detection};
\node [left of=det,node distance=5em] (output) {$\widehat{\mathbf{s}}$};
\draw[dashed] ($ (iq1) - (1.2,-1) $)  rectangle ($ (noise) + (1.2,-1) $)  node[midway,yshift=3.5em]{Physical Channel Model};
\draw[dashed] ($ (det) - (1.2,-1) $)  rectangle ($ (iq2r) + (1.2,-1) $)  node[midway,yshift=-3.5em]{PhyCOM};
\draw[->,>=latex] (start) -- (mod);
\draw[->,>=latex] (mod) -- (iq1);
\draw[->,>=latex] (iq1) -- (pn1);
\draw[->,>=latex] (pn1) -- (fir);
\draw[->,>=latex] (fir) -- (pn2);
\draw[->,>=latex] (pn2) -- (iq2);
\draw[->,>=latex] (iq2) -- (noise);
\draw[->,>=latex] (noise)--++(1.5,0) |- (iq2r) node[below,xshift=10em] {$\tilde{\mathbf{y}}_0$};
\draw[->,>=latex] (iq2r) -- (pn2r);
\draw[->,>=latex] (pn2r) -- (firr);
\draw[->,>=latex] (firr) -- (pn1r);
\draw[->,>=latex] (pn1r) -- (iq1r);
\draw[->,>=latex] (iq1r) -- (det);
\draw[->,>=latex] (det) -- (output);
\end{tikzpicture}
\caption{Communication Model with PhyCOM model mismatch. The physical layer is composed of a transmitter with IQ impairment ($\boldsymbol \theta_1=[0.9,0.4,-0.4,0.6]^T$) and Wiener phase noise, a FIR channel composed of $8$ taps ($\mathbf{h}=[0.9+0.1j,0.3+0.3j,0.1+0.05j,0.02+0.1j,0.1-0.05j,0.02-0.1j,0.1+0.03j,0.04-0.012j]^T$), and a receiver with IQ impairment ($\boldsymbol \theta=[1.8,0.1,0.13,0.8]^T$) and Wiener phase noise. The noise layer add a circular white Gaussian noise $\mathcal{N}_{\mathcal{C}}(0,\sigma^2)$.}\label{fig:channel2}
\end{figure*}

In the next simulation, we consider a more complex channel communication model with time-varying parameters that is sometimes encountered in coherent optical communications. The communication model is described in Fig.~\ref{fig:channel2}. This model is composed of a transmitter corrupted by IQ imbalance and phase noise, an FIR channel, and a receiver corrupted by IQ imbalance and phase noise. Statistically, the phase noise is generated using a discrete Wiener process
\begin{align}
\varphi[n] = \varphi[n-1]+b[n],
\end{align}
where $b[n]\sim \mathcal{N}(0,\sigma^2_b)$ with $\sigma^2_b=2\pi (5\times 10^{-5})$. In each simulation, the phase noise realizations at the transmitter and receiver sides are generated using independent Wiener processes. To our knowledge, there is no standard DSP technique able to tackle this complex compensation problem. While the compensation of this particular channel model with conventional DSP techniques is still an open issue, the flexibility of the proposed PhyCOM architecture allows to easily compensate for all channel impairments.
For impairments compensation, we propose to use a PhyCOM network composed of 6 layers. To track the phase noise, the PhyCOM network includes two Quasi-Static (QS) phase noise layers (see Fig.~\ref{fig:channel2}). Note that this compensation strategy leads to a mismatch between the true communication model and the PhyCOM model. The training symbols are generated using a pilot-based strategy where the $N_p$ pilots are inserted uniformly in the data block (see Fig.~\ref{fig_allocate}). The QS phase noise layers are initialized with zero vectors of length $K_p$. The other layer parameters are initialized with the same values as in the previous scenario. Finally, the PhyCOM network is parametrized by $K=24+2K_p$ real-valued parameters. In the following simulations, we only report on the performance of the PhyCOM1 and PhyCOM2 techniques since, to our knowledge, there is no standard DSP technique for the compensation of this complex channel model.

\pgfplotscreateplotcyclelist{mark list2}{%
    every mark/.append style={solid,fill=black},mark=o,only marks,thick\\%
    every mark/.append style={solid,fill=black},mark=none,thick,black\\%
    every mark/.append style={solid,fill=blue},mark=none,thick,blue,dashed\\%
    every mark/.append style={solid,fill=blue},mark=star,thick,blue,solid\\%
    every mark/.append style={solid,fill=blue},mark=none,thick,brown,dashed\\%
  every mark/.append style={solid,fill=blue},mark=diamond,brown,thick,solid\\%
    every mark/.append style={solid,fill=red},mark=none,thick,red,dashed\\%
    every mark/.append style={solid,fill=red},mark=+,thick,red,solid\\%
}

\begin{figure}[!t]
\centering
\begin{subfigure}{.48\textwidth}
\begin{tikzpicture}
    \begin{semilogyaxis}[width=1.\textwidth,xlabel near ticks,ylabel near ticks,grid=both,ylabel=MSE,xlabel=SNR (dB),xmin=0,ymax=1,xmax=40,ymin=0.001,legend pos=south west,cycle list name=mark list2,legend style={font=\footnotesize}]
    	\addplot table[x index=0, y=theo, col sep=comma] {./csv/simulation4/mse_test.csv}; 				
    	\addplot table[x index=0, y=clairvoyant, col sep=comma] {./csv/simulation4/mse_test.csv}; 				
      	\addplot  table[x index=0, y=phycom1_0, col sep=comma] {./csv/simulation4/mse_test.csv}; 				
 	\addplot table[x index=0, y=phycom2_0, col sep=comma] {./csv/simulation4/mse_test.csv}; 		
     	\addplot table[x index=0, y=phycom1_5, col sep=comma] {./csv/simulation4/mse_test.csv}; 			
	\addplot table[x index=0, y=phycom2_5, col sep=comma] {./csv/simulation4/mse_test.csv}; 
	\addplot table[x index=0, y=phycom1_10, col sep=comma] {./csv/simulation4/mse_test.csv}; 
	\addplot table[x index=0, y=phycom2_10, col sep=comma] {./csv/simulation4/mse_test.csv}; 
	 \legend{\scriptsize Theoretical, \scriptsize Clairvoyant,\scriptsize  PhyCOM1 $K_p=0$,\scriptsize  PhyCOM2 $K_p=0$,\scriptsize  PhyCOM1 $K_p=5$, \scriptsize PhyCOM2 $K_p=5$,\scriptsize  PhyCOM1 $K_p=10$,\scriptsize  PhyCOM2 $K_p=10$}
    \end{semilogyaxis}
   \end{tikzpicture}
\caption{Evolution of the MSE.}\label{mse2_fig1}
\end{subfigure}
\begin{subfigure}{.48\textwidth}
\centering
\begin{tikzpicture}
    \begin{semilogyaxis}[width=1.0\textwidth,xlabel near ticks,ylabel near ticks,grid=both,ylabel=SER,ymin=0.001,xmin=0,xmax=40,xlabel=SNR (dB),legend pos=south west,cycle list name=mark list2,legend style={font=\footnotesize},cycle list shift=1]
    	\addplot table[x index=0, y=clairvoyant, col sep=comma] {./csv/simulation4/ser_test.csv}; 				
      	\addplot  table[x index=0, y=phycom1_0, col sep=comma] {./csv/simulation4/ser_test.csv}; 				
 	\addplot table[x index=0, y=phycom2_0, col sep=comma] {./csv/simulation4/ser_test.csv}; 		
     	\addplot table[x index=0, y=phycom1_5, col sep=comma] {./csv/simulation4/ser_test.csv}; 			
	\addplot table[x index=0, y=phycom2_5, col sep=comma] {./csv/simulation4/ser_test.csv}; 
	\addplot table[x index=0, y=phycom1_10, col sep=comma] {./csv/simulation4/ser_test.csv}; 
	\addplot table[x index=0, y=phycom2_10, col sep=comma] {./csv/simulation4/ser_test.csv}; 
     \legend{\scriptsize Clairvoyant,\scriptsize  PhyCOM1 $K_p=0$,\scriptsize  PhyCOM2 $K_p=0$,\scriptsize  PhyCOM1 $K_p=5$,\scriptsize  PhyCOM2 $K_p=5$,\scriptsize  PhyCOM1 $K_p=10$,\scriptsize  PhyCOM2 $K_p=10$}
    \end{semilogyaxis}
\end{tikzpicture}
\caption{Evolution of the SER.}\label{mse2_fig2}
\end{subfigure}
\caption{Evolution of the errors versus SNR ($N_p=50$ symbols).}\label{ser2fig}
\end{figure}
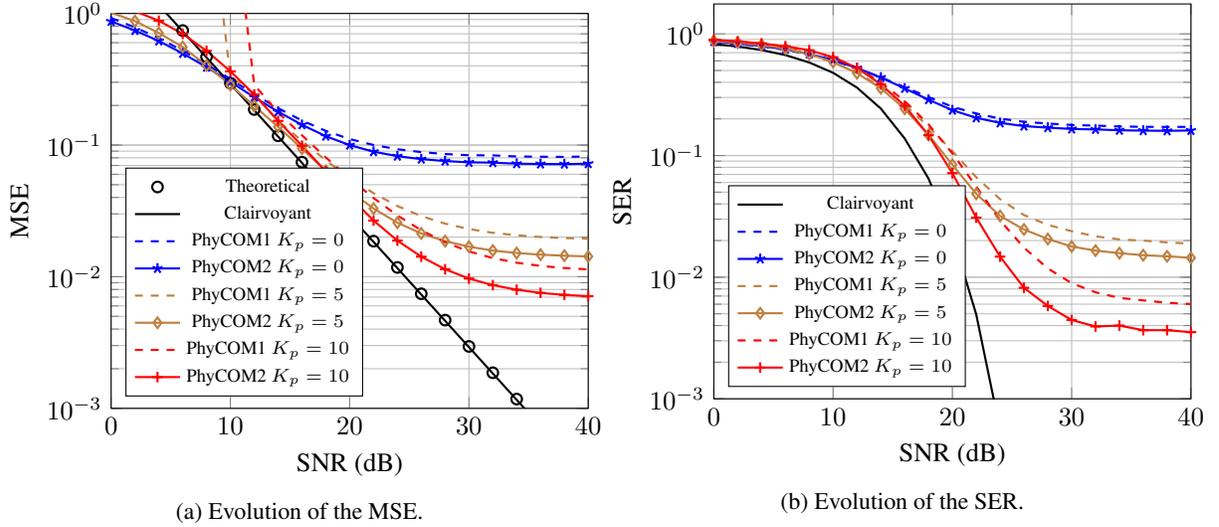

\subsubsection{MSE Analysis}

Fig.~\ref{mse2_fig1} shows the evolution of the MSE for the clairvoyant PhyCOM with perfect compensation of the phase noise, and 3 PhyCOM networks with $K_p=0$ (no quasi-static phase noise layer), $K_p=5$ and $K_p=10$, respectively. We note that the PhyCOM networks without phase noise compensation ($K_p=0$) do not provide a reliable estimate of the transmitted data since the MSE is still larger than $0.07$ at large SNRs. Concerning the PhyCOM networks with $K_p=5$ and $K_p=10$, we observe that the first one performs slightly better than the second one for SNR below $16$dB and that this trend seems reversed for larger SNRs. At SNR$=40$ dB, the MSE is equal to $0.014$ and $0.007$ for $K_p=5$ and $K_p=10$, respectively. This behavior illustrates that setting an optimal value of $K_p$ can be a difficult task since the optimal value depends on the SNR.

\subsubsection{SER Analysis}

Fig.~\ref{mse2_fig2} presents the influence of the SNR on the estimation and detection performances. As previously observed, we note that the PhyCOM networks without phase noise compensation ($K_p=0$) yield poor performance even at large SNR (SER$\approx0.16$ for the testing data). The introduction of transmitter and receiver quasi-static phase noise layers significantly reduces the SER. More precisely, the SER is equal to $0.014$ and $0.0035$ for $K_p=5$ and $K_p=10$, respectively. For this particular problem, additional simulations have shown that larger values of $K_p$ do not necessarily improve the SER (SER $=0.004$ for $K_p=20$ and SNR=$40$dB) and tend to increase the generalization error since the number of networks parameters becomes too large in comparison to the number of pilot symbols.

\section{Conclusion}
This paper describes a new multi-layer network, called PhyCOM, for linear impairments compensation and detection in communication systems. The PhyCOM network combines both the benefits of parametric techniques and the flexibility of feedforward networks. The structure of the PhyCOM network is composed of widely linear layers parametrized by a small number of unknown parameters, and can be trained using a semi-supervised strategy with a small number of pilot symbols.

Simulation results have shown that the proposed network can compensate a wide range of SISO impairment layers, such as IQ imbalance, FIR channel, carrier frequency offset, and phase noise, with a small number of training symbols ($\approx 50$ symbols in most simulations). While these results are promising, we believe that performance improvement could be obtained by tracking the noise distribution through the layers, by developing more subtle semi-supervised training algorithms, or by plugging deep-learning detectors at the end of the proposed network. Furthermore, we are convinced that the proposed technique can be expanded to broader applications by including MIMO, nonlinear, and stochastic layers.

\bibliographystyle{IEEEtran}  
\bibliography{biblio/biblio.bib}

\end{document}